\documentclass[aps,prd,preprint,groupedaddress]{revtex4-1}
\usepackage{amsmath,amssymb}
\usepackage{graphicx}

\begin{document}

\preprint{KUNS-2504}

\title{Bimetric gravity and two-component fluid in the AdS/CFT correspondence}


\author{Kouichi Nomura}
\affiliation{Department of Physics, Kyoto University, Kyoto, 606-8502, Japan}


\date{\today}

\begin{abstract}
We study bimetric gravity through the context of the AdS/CFT correspondence, especially, in the first order hdrodynamic limit. 
If we put pure general relativity as a bulk field, the boundary field theory is interpreted as fluid of the $\mathcal{N}=4$ supersymmetric Yang-Mills plasma. 
The transport coefficients of this plasma are computed via the AdS/CFT correspondence. 
Then, we prepare a pair of gravitational fields on the bulk side and let them interact. 
We expect that two-component fluid emerge on the CFT boundary side because the number of metrics becomes double. 
However, the situation is rather complicated. 
The interaction generates a massive graviton. This massive mode leads to the extra divergences which are absent in the case of general relativity. 
Our first investigation is how to cancel these divergences. 
After that, we see the emergence of two-component fluid and calculate their pressure and sheer viscosity. 
The interaction makes two kinds of fluid mixed and the sheer viscosity obtain slight correction dependent on the mass of a graviton. 
\end{abstract}

\pacs{}

\maketitle

\section{Introduction}
The AdS/CFT correspondence is one of the most widely studied topics in modern theoretical physics \cite{Maldacena:1997re, Gubser:1998bc, Witten:1998qj, Aharony:1999ti}. 
It covers string theory, general relativity, condensed matter physics and so on. 
It relates a gravity theory on a (d+1)-dimensional asymptotically AdS space-time to some matter field theory on the d-dimensional boundary. When the bulk side is weakly coupled, 
the coupling of the boundary field gets strong. Therfore, we can investigate complicated matter field theories through the calculation of rather simple equations from bulk gravity theories. 
Applications of the AdS/CFT correspondence are varied, for example, supersymmetric Yang-Mills plasma as a clue to quark-gluon plasma, superconductor, non-fermi liquid and so on \cite{Policastro:2001yc, Horowitz:2010gk, Iqbal:2011ae}. \par
In the standard settings of the AdS/CFT correspondence, we put pure general relativity on the bulk side with other fields (scalar, vector, spinor, etc). 
Sometimes, massive fields play an important role, for instance, we use a massive scalar in holographic superconductor. Therefore, a question arises what if we put 
a theory of massive gravity instead of massless general relativity. The effect of massive gravitons on the AdS/CFT correspondence has been asked several times in the past \cite{Kiritsis:2006hy, Aharony:2006hz, Apolo:2012gg}. 
They say when two or more CFT boundaries are prepared and their interaction is swiched on, some gravitons on the bulk side get massive. 
This situation makes us remember a theory of bimetric or multimetric gravity. 
Generally, theories of bimetric or multimetric gravity has only one overall diffeomorphism invariance. Hence, only one graviton remains massless and others get massive. 
Interaction of the gravitational fields makes massive gravitons. 
Here, we reverse the picture and ask what if we put interacting gravitational fields on the bulk side. 
Until very recently, no consistent theory of interacting gravitational fields had been known. They had suffered from emergence of the extra ghost degrees of freedom. 
However, the ghost problem has been overcome \cite{Hassan:2011tf, Hassan:2011ea} and we now have the consistent theories of bimetric or multimetric gravity \cite{Hassan:2011zd, Hinterbichler:2012cn, Nomura:2012xr}. 
Hence, in this paper, we attempt to consider bimetric gravity on the context of the AdS/CFT correspondence. \par
In studying the AdS/CFT correspondence, one of the main difficulties is how to interpret the result. We put some gravity theory on the bulk asymptotically AdS space-time, and calculate 
correlation functions of the boundary field theory. However, we cannot know in advance what kind of theory we have on the CFT boundary. 
Sometimes, we are puzzled by a question what is the physical meaning. 
Therefore, we proceed as close as possible to the well-known case where pure general relativity is used.  Besides, we rely on the hydrodynamic limit, which makes analytic calculation possible. 
In these settings, the counterpart on the boundary side is interpreted as fluid of the supersymmetric Yang-Mills plasma \cite{Policastro:2002se, Policastro:2002tn}, where the transport coefficients such as 
sheer viscosity are calculated. Following this, we investigate the case of bimetric gravity and see that two-component fluid emerges. We also calculate the values of their pressure and sheer viscosity. \par
The organization of this paper is as follows. In section \ref{massivegravity}, we review the case of pure general relativity and apply the method to dRGT massive garavity. 
Bimetric gravity is a genaralization of dRGT massive gravity. Therefore, in section \ref{bimetricgravity}, we further extend the previous result. Section \ref{conclusion} is devoted to the 
conclusion.

\section{dRGT massive gravity and the AdS/CFT correspondence in the first order hydrodynamic limit}\label{massivegravity}
In this section, we consider the AdS/CFT correspondence in dRGT massive gravity \cite{deRham:2010ik, deRham:2010kj, Hassan:2011vm, Hassan:2011tf, Hassan:2011ea}. 
We focus on the first order hydrodynamic limit where we can easily carry out the analytic calculation. 
To begin with, we review the case of general relativity \cite{Policastro:2001yc, Policastro:2002se} (See a review \cite{Natsuume} for the detailed calculation.) 
and extend it to that of massive gravity. In the calculation of the AdS/CFT correspondence in general relativity, we encounter divergent terms and add a counterterm to cancel them. 
We see that mass of a graviton gives rise to extra divergences which are absent in the case of general relativity. 
The main topic of this section is how to cancel these additional divergences.

\subsection{the case of general relativity}
In this subsection, we revisit the AdS/CFT correspondence in general relativity. 
When we take a long wave-length limit, the matter field theory on the CFT boundary is considered as the supersymmetric Yang-Mills plasma. 
The pressure and the sheer viscosity of this plasma can be calculated through the AdS/CFT corresponcence. 
In this paper, we take only the first order hydrodynamic limit which is enough to obtain the values of the pressure and the viscosity. 
The hydrodynamic limit means a long wave-length limit, and we carry out the calculation only up to the first order derivative expansion. \par 
We start the case of general relativity with the following action 
\begin{align}
&S=S_{EH}+S_{GH}+S_{ct}\\
&S_{EH}=\frac{1}{16\pi G_5}\int d^5x \sqrt{-g} \big(R-2\Lambda) \\
&S_{GH}=\frac{2}{16\pi G_5}\int_{AdS-bdy} d^4x \sqrt{-\gamma}K\\
&S_{ct}=\frac{1}{16\pi G_5}\int_{AdS-bdy} d^4x \sqrt{-\gamma}\Big(\frac{6}{L}+\frac{L}{2}\mathcal{R}+\cdots \Big),
\end{align}
where we consider a five dimensional asymptotically AdS space-time and $AdS-bdy$ stands for the AdS-boundary. The first term $S_{EH}$ is Einstein-Hilbert action, the second $S_{GH}$ is 
Gibbons-Hawking term and the third $S_{ct}$ is a counterterm added to cancel divergences. 
$S_{EH}$ is a bulk term, and boundary terms are $S_{GH}$ and $S_{ct}$. 
The metric $\gamma$ is a four dimensional induced metric on the AdS-boundary and $K$ is the extrinsic curvature. $\mathcal{R}$ is the curvature constructed from $\gamma$. 
$L$ is the AdS-radius and related to the cosmological constant as $\Lambda=-6/L^2$. 
In the counterterm $S_{ct}$, we neglected higher order derivative terms such as $\mathcal{R}^2$ because they are not needed in the first order hydrodynamic limit. 
(In fact, we do not need $\mathcal{R}$, too.) 
The hydrodynamic limit 
is a long wave-length limit, and especially we focus on the first order limit. We neglect terms containing higher than second order derivatives with respect to 
the coordinates on the AdS-boundary. \par
In general relativity, a lot of asymptotically AdS solutions are known. However, in this paper, we focus only on five dimensional Schwarzschild AdS black hole (SAdS-BH). 
This metric is given by
\begin{align}
g_{\mu \nu}dx^{\mu}dx^{\nu}&=\Big(\frac{r_0}{L}\Big)^2\frac{1}{u^2}\big(-h dt^2+dx^2+dy^2+dz^2 \big)+\frac{L^2}{hu^2}du^2, \label{metAdSBH}
\end{align}
where $r_0$ is a constant and $L$ is the AdS-radius. We set the coodinates as $x^\mu=(t,x,y,x,u)$. 
$h$ is defined as $h=1-u^4$ ($0<u<1$). The AdS-boundary is located at $u=0$  and the Black Hole horizon is on the region $u=0$. 
If we set $h=1$ and $r_0=1$, we have pure AdS space-time. \par
According to the ordinary prescription of the AdS/CFT correspondence, we consider perturbations around SAdS-BH and expand the action up to the second order. 
Then, we solve the equation of motion, and substitute the solution back into the action. To get this on-shell action is the first step.  
For simplicity, we take a perturbation $g_{\mu \nu}=\bar{g}_{\mu \nu}+\delta g_{\mu \nu}$ (background +fluctuation) such as 
\begin{align}
{\delta g^{\mu}}_{\nu}=\bar{g}^{\mu \lambda}\delta g_{\lambda \nu}=\left(
\begin{array}{ccccc}
0 & 0& 0& 0& 0 \\
0 & 0& \phi&0 & 0 \\
0 & \phi& 0& 0&0  \\
0 & 0& 0& 0& 0 \\
0 & 0& 0& 0& 0
\end{array}
\right)  ,\qquad \phi=\phi(t,u). \label{spur}
\end{align}
Here, we assume that $\phi$ depends only on $t$ and $u$. Then, we expand the action $S=S_{EH}+S_{GH}+S_{ct}$ up to the second order in $\phi$ and perform the Fourier transform $\phi(t,u)=\int \frac{d\omega}{2\pi} e^{-i\omega t}\phi_\omega(u)$. \par
To begin with, we set about perturbed Einstein-Hilbert action
\begin{align}
S_{EH}&=-\frac{V_4}{16\pi G_5}\frac{r_0^4}{L^5}\int^1_0 du \frac{8}{u^5} \\
      &+\frac{V_3}{16\pi G_5}\frac{r_0^4}{L^5}\int \frac{d\omega}{2\pi}\int^1_0 du \bigg\{\frac{3}{2}\frac{h}{u^3}\phi'_{-\omega} \phi'_{\omega}+2\frac{h}{u^3}\phi_{-\omega} \phi''_\omega-\frac{8}{u^4}\phi_{-\omega} \phi'_{\omega}+\bigg(\frac{1}{2u^3h}\Big(\frac{L^2}{r_0}\omega \Big)^2+\frac{4}{u^5}\bigg)\phi_{-\omega}\phi_{\omega}\bigg\} \label{SEH},
\end{align}
where we abbreviated $\int dxdydz=V_3$, $\partial_u \phi(u)=\phi'$ and $\int dtdxdydz=V_4$. We cannot discard total derivatives because we have boundaries. \par
If we write the bulk action as 
\begin{align}
S_{bulk}=S_{EH}=\int^1_0 du \, \mathcal{L}(\phi \, ,\, \phi' \, ,\, \phi'') \label{Sbulk}
\end{align}
and take a variation $\phi \rightarrow \phi+\delta\phi$, we obtain the variation of the bulk action
\begin{align}
\delta S_{bulk}=& \int^1_0 du \, \bigg\{\Big(\frac{\partial \mathcal{L}}{\partial \phi}\Big)\delta \phi+\Big(\frac{\partial \mathcal{L}}{\partial \phi'}\Big)\delta \phi'+\Big(\frac{\partial \mathcal{L}}{\partial \phi''}\Big)\delta \phi''\bigg\} \\
        =& \bigg[ \bigg\{ \Big(\frac{\partial \mathcal{L}}{\partial \phi'}\Big)-\Big(\frac{\partial \mathcal{L}}{\partial \phi''}\Big)'\bigg\}\delta \phi +\Big(\frac{\partial \mathcal{L}}{\partial \phi''}\Big)\delta \phi'\bigg]_0^1
          +\int^1_0 du \, \bigg\{\Big(\frac{\partial \mathcal{L}}{\partial \phi''}\Big)''-\Big(\frac{\partial \mathcal{L}}{\partial \phi'}\Big)'+\Big(\frac{\partial \mathcal{L}}{\partial \phi}\Big) \bigg\}\delta \phi.
\end{align}
The $\delta \phi'$ term will be canceled by $S_{GH}$, and we get the equation of motion (EOM)
\begin{align}
\text{EOM : }\Big(\frac{\partial \mathcal{L}}{\partial \phi''}\Big)''-\Big(\frac{\partial \mathcal{L}}{\partial \phi'}\Big)'+\Big(\frac{\partial \mathcal{L}}{\partial \phi}\Big)=0.
\end{align}
While, the Lagrangian density $\mathcal{L}$ contains the zeroth and the secod order terms in $\phi$ (see Eq.(\ref{SEH})). We can write 
\begin{align}
S_{bulk}=& -\frac{V_4}{16\pi G_5}\frac{r_0^4}{L^5}\int_0^1 du \frac{8}{u^5}+\frac{1}{2} \int_0^1 du \, \bigg\{\Big(\frac{\partial \mathcal{L}}{\partial \phi}\Big) \phi+\Big(\frac{\partial \mathcal{L}}{\partial \phi'}\Big) \phi'+\Big(\frac{\partial \mathcal{L}}{\partial \phi''}\Big) \phi''\bigg\} \\
        =& \frac{V_4}{16\pi G_5}\frac{r_0^4}{L^5}\Big[\frac{2}{u^4}\Big]_0^1+\frac{1}{2}\bigg[ \bigg\{ \Big(\frac{\partial \mathcal{L}}{\partial \phi'}\Big)-\Big(\frac{\partial \mathcal{L}}{\partial \phi''}\Big)'\bigg\} \phi +\Big(\frac{\partial \mathcal{L}}{\partial \phi''}\Big) \phi'\bigg]_0^1\\
          &+\frac{1}{2}\int_0^1 du \, \bigg\{\Big(\frac{\partial \mathcal{L}}{\partial \phi''}\Big)''-\Big(\frac{\partial \mathcal{L}}{\partial \phi'}\Big)'+\Big(\frac{\partial \mathcal{L}}{\partial \phi}\Big) \bigg\} \phi.
\end{align}
According to the AdS/CFT prescription, we solve the EOM and substitute the solution to the original action, and get the on-shell action. 
Thus, the last term is discarded. Besides, we do not need terms coming from the field value $\phi(u=1)$ on the non AdS-boundary \cite{Son:2002sd}, so they are neglected. Then, the bulk action is read as 
\begin{align}
S_{bulk}=\frac{V_4}{16\pi G_5}\frac{r_0^4}{L^5}\Big(2-\frac{2}{u^4}\Big)\Big|_{u=0} -\frac{1}{2}\bigg\{ \bigg( \Big(\frac{\partial \mathcal{L}}{\partial \phi'}\Big)-\Big(\frac{\partial \mathcal{L}}{\partial \phi''}\Big)'\bigg) \phi +\Big(\frac{\partial \mathcal{L}}{\partial \phi''}\Big) \phi'\bigg\}\Bigg|_{u=0}. \label{Sbulkbd}
\end{align} \par
In the case of general relativity, $S_{bulk}=\int du \, \mathcal{L}(\phi \, ,\, \phi' \, ,\, \phi'')$ is given by Eq.(\ref{SEH}). 
Therefore, we obtain the explicit formulae of the EOM and the bulk action $S_{bulk}$
\begin{align}
\text{EOM : } & \Big(\frac{h}{u^3}\phi'_\omega \Big)'+\Big(\frac{L^2}{r_0}\omega \Big)^2 \frac{1}{u^3h}\phi_\omega=0 \label{EOM}\\
S_{bulk}=& \frac{V_4}{16\pi G_5}\frac{r_0^4}{L^5}\Big(2-\frac{2}{u^4}\Big)+\frac{V_3}{16\pi G_5}\frac{r_0^4}{L^5}\int \frac{d\omega}{2\pi}\bigg\{\frac{h}{u^4}\phi_{-\omega}\phi_\omega-\frac{3}{2}\frac{h}{u^3}\phi_{-\omega} \phi'_\omega \bigg\}\Bigg|_{u=0} \label{hSbulk} \\ 
                =& \frac{V_4}{16\pi G_5}\frac{r_0^4}{L^5}\Big(2-\frac{2}{u^4}\Big)+\frac{V_3}{16\pi G_5}\frac{r_0^4}{L^5}\int \frac{d\omega}{2\pi}\bigg\{\Big(\frac{1}{u^4}-1\Big)\phi_{-\omega}\phi_\omega-\frac{3}{2}\Big(\frac{1}{u^3}-u\Big)\phi_{-\omega} \phi'_{\omega}\bigg\}\Bigg|_{u=0}.\label{HSbulk}
\end{align}
The first term in $S_{bulk}$ is divergent as $\frac{1}{u^4}|_{u=0}$. $S_{ct}$ is added to cancel this divergence. \par
Next, we continue to calculate the boundary terms $S_{GH}$ and $S_{ct}$.
Using $\sqrt{-\gamma}\sim \big(\frac{r_0}{L}\big)^4\frac{\sqrt{h}}{u^4}\big(1-\frac{1}{2}\phi^2\big)$ and $K=-\frac{\partial_u \sqrt{\gamma}}{N\sqrt{\gamma}}$ with a lapse $N^{-1}=u\sqrt{h}/L$, 
the expansion of $S_{GH}$ up to the second order in $\phi$ is given by 
\begin{align}
S_{GH}&=-\frac{2}{16\pi G_5}\int d^4x\frac{1}{N}\partial_{u}\sqrt{-\gamma}\Big|_{u=0} \\
      &=\frac{1}{16\pi G_5}\frac{r_0^4}{L^5}\int d^4x\bigg\{8\frac{h}{u^4}-\frac{h'}{u^3}+\Big(\frac{h'}{2u^3}-4\frac{h}{u^4}\Big)\phi \phi+2\frac{h}{u^3}\phi \phi'\bigg\} \Big|_{u=0} \label{hSGH} \\
      &=\frac{V_4}{16\pi G_5}\frac{r_0^4}{L^5}\Big(\frac{8}{u^4}-4\Big)+\frac{V_3}{16\pi G_5}\frac{r_0^4}{L^5}\int \frac{d\omega}{2\pi}\bigg\{\Big(-\frac{4}{u^4}+2\Big)\phi_{-\omega}\phi_{\omega}+2\Big(\frac{1}{u^3}-u\Big)\phi_{-\omega}\phi'_{\omega}\bigg\}\Bigg|_{u=0}.
\end{align}
The curvature $\mathcal{R}$ contains derivatives higher than second order and its background value is zero. Thus, we do not need it in the first order hydrodynamic limit, and we obtain
\begin{align}
S_{ct}=&-\frac{1}{16\pi G_5}\int d^4x\sqrt{-\gamma}\frac{6}{L}\Big|_{u=0}\\
      =&-6\frac{V_4}{16\pi G_5}\frac{r_0^4}{L^5}\frac{\sqrt{h}}{u^4}+\frac{V_3}{16\pi G_5}\frac{r_0^4}{L^5}\int \frac{d\omega}{2\pi}3\frac{\sqrt{h}}{u^4}\phi_{-\omega}\phi_\omega \Big|_{u=0} \label{hSct} \\
      =&\frac{V_4}{16\pi G_5}\frac{r_0^4}{L^5}\Big(3-\frac{6}{u^4}\Big)+\frac{V_3}{16\pi G_5}\frac{r_0^4}{L^5}\int \frac{d\omega}{2\pi}\Big(\frac{3}{u^4}-\frac{3}{2}+O[u^4]\Big)\phi_{-\omega}\phi_\omega\Big|_{u=0},
\end{align}
where $O[u^4]$ stands for higher order terms than $u^4$. 
Then, $S_{ct}$ cancels the zeroth order divergent term and we obtain
\begin{align}
S&=S_{bulk}+S_{GH}+S_{ct}\\
 &=\frac{V_4}{16\pi G_5}\frac{r_0^4}{L^5}+\frac{V_3}{16\pi G_5}\frac{r_0^4}{L^5}\int \frac{d\omega}{2\pi}\bigg\{ \Big(-\frac{1}{2}+O[u^4]\Big)\phi_{-\omega} \phi_{\omega} +\frac{1}{2}\Big(\frac{1}{u^3}-u\Big)\phi_{-\omega}\phi'_{\omega}\bigg\} \Bigg|_{u=0}. \label{Sren}
\end{align} 
We put a derivative on $\phi_{\omega}$ in order to follow the Minkowski prescription \cite{Son:2002sd}. \par
Now, we solve the EOM Eq.(\ref{EOM}) and substitute the solution into Eq.(\ref{Sren}) to get the on-shell action.
Because our main interest of this paper is the first order hydrodynamic limit, we have only to solve the EOM up to the first order expansion in $\omega$. 
We expand $\phi_\omega(u)$ as 
\begin{align}
\phi_\omega(u)=\phi_0(u)+\omega \phi_1(u)+\omega^2 \phi_2(u)+\cdots 
\end{align}
and insert it into the EOM Eq.(\ref{EOM})
\begin{align}
\Big(\frac{h}{u^3}\phi'_i \Big)'=0\qquad (i=0,1).
\end{align}
The solution is written as 
\begin{align}
\phi_i=A_i+B_i\ln(1-u^4) \qquad (i=0,1).
\end{align}
Thus, we obtain 
\begin{align}
\phi_{\omega}(u)=(A_0+\omega A_1)+(B_0+\omega B_1)\ln(1-u^4)+O[\omega^2] \label{ep},
\end{align}
where $A_i$ and $B_i$ are constants. 
This solution is substituted into Eq.(\ref{Sren}), but in Eq.(\ref{Sren}) we need only the asymptotic formula near the AdS-boundary $u\sim 0$. 
We can write, abbreviated as $A_{\omega}=A_0+\omega A_1$ and $B_{\omega}=B_0+\omega B_1$,
\begin {align}
\phi_\omega(u)=A_\omega+B_\omega\ln(1-u^4)+O[\omega^2] = A_{\omega}+B_{\omega}(-u^4+O[u^8])+O[\omega^2].
\end{align}
Then, the substituted on-shell action is obtained
\begin{align}
S=\frac{V_4}{16\pi G_5}\frac{r_0^4}{L^5}+\frac{V_3}{16\pi G_5}\frac{r_0^4}{L^5}\int \frac{d\omega}{2\pi}\Big(-\frac{1}{2}A_{-\omega}A_\omega -2A_{-\omega}B_\omega \Big) \label{Ssub}.
\end{align}

The remainig constants $A_{0,1}$ and $B_{0,1}$ are fixed from the boundary condition on the Black Hole horizon ($u=1$), so we have to solve the EOM Eq.(\ref{EOM}) near the $u\sim 1$ region. 
Approximating as $h=1-u^4\sim 4(1-u)$, we get the near horizon EOM from Eq.(\ref{EOM})
\begin{align}
\phi_{\omega}''-\frac{1}{1-u}\phi'_{\omega}+\Big(\frac{L^2}{4r_0}\omega \Big)^2\frac{1}{(1-u)^2}\phi_{\omega}=0,
\end{align}
and the solution is given by 
\begin{align}
\phi_{\omega}(u\sim1)\propto (1-u)^{\pm i\frac{L^2}{4r_0}\omega}.
\end{align}
Here, we set $r_0=L=1$ and remember that background SAdS-BH has the metric $ds^2=-\frac{h}{u^2}dt^2+\frac{du^2}{u^2h}$. 
If we change the coordinate $u$ to $u_{*}=\int^{\infty}_u\frac{du}{h}\sim \frac{1}{4}\ln(1-u)$, this meric reads as $ds^2\propto (-dt^2+du^2)$. 
Then, the solution is written as $\phi_{\omega}(u\sim1)\propto (1-u)^{\pm \frac{i}{4}\omega}=e^{\pm i\omega u_*}$, and we have 
$\phi(t,u\sim 1)\propto e^{-i\omega t}\phi_{\omega}(u\sim 1)=e^{-i\omega(t\mp u_*)}$. Because the Black Hole horizon is located at $u_*=-\infty$, 
the solution $e^{-i\omega(t+ u_*)}$ represents an ingoing wave and the other $e^{+i\omega(t+ u_*)}$ is outgoing. 
We select the ingoing wave condition according to the sandard AdS/CFT prescription. Thus, we obtain the near horizon solution
\begin{align}
\phi_\omega(u\sim1)\propto (1-u)^{-i\frac{L^2}{4r_0}\omega}\sim (1-u^4)e^{-i\frac{L^2}{4r_0}\omega}= 1-i \frac{L^2}{4r_0}\omega \ln(1-u^4)+O[\omega^2]. \label{np}
\end{align}
The previously obtained solution Eq.(\ref{ep}) must match Eq.(\ref{np}), which fixes the constants as $A_1=0$, $B_0=0$, $B_1=-i\frac{L^2}{4r_0}A_0$.
Then, renaming $A_0$ as $\phi^{(0)}$, we obtain the on-shell action from Eq.(\ref{Ssub})
\begin{align}
S=\frac{V_4}{16\pi G_5}\frac{r_0^4}{L^5}+\frac{V_3}{16\pi G_5}\frac{r_0^4}{L^5}\int \frac{d\omega}{2\pi}\bigg\{-\frac{1}{2}\phi^{(0)}_{-\omega}\phi_\omega^{(0)} +\frac{1}{2}\phi^{(0)}_{-\omega}\Big(i\frac{L^2}{r_0}\omega \Big)\phi_\omega^{(0)}\bigg\},
\end{align} 
where we should notice that $i\int d\omega \phi^{(0)}_{-\omega}\omega \phi^{(0)}_{\omega}$ cannot be interpreted as zero \cite{Son:2002sd}. \par
The last step is to apply the GKP-Witten relation 
\begin{align}
\Big \langle  \exp \Big(i \int \phi^{(0)}\mathcal{O}\Big)   \Big \rangle =\exp \Big(iS\big[\phi|_{u=0}=\phi^{(0)}\big]\Big). \label{GKPW}
\end{align}
In the right hand side, we have the action $S[\phi]$. We solve the EOM of the bulk field $\phi$, where we denote the boundary value of the solution as $\phi^{(0)}$. 
The solution of the EOM is substituted into the action and we get the on-shell action $S\big[\phi|_{u=0}=\phi^{(0)}]$. 
The left hand side represents the expectation value of the boundary field theory, where $\phi^{(0)}$ becomes a source of an operator $\mathcal{O}$. \par
In our case, $\phi$ is a fluctuation of the spin-2 field so that $\mathcal{O}$ is interpreted as a perturbed boundary energy-momentum tensor $\delta T_{\mu \nu}$. Thus, we obtain 
\begin{align}
<\delta T^{xy}_{\omega}>=\frac{\delta S}{\delta \phi^{(0)}_{-\omega}}=-\frac{1}{16\pi G_5}\frac{r_0^4}{L^5}\phi^{(0)}_{\omega}+i\frac{1}{16\pi G_5}\Big(\frac{r_0}{L}\Big)^3\omega \phi^{(0)}_{\omega},\label{EMAds}
\end{align}
where the functional derivative is interpreted as $\frac{\delta}{\delta \phi_{-\omega}}\phi_{-\omega}\mathcal{F}_{\omega}\phi_{\omega}=2\mathcal{F}_{\omega}\phi_\omega$ \cite{Son:2002sd}. 
We neglected $V_3$, because $V_3$ is interpreted as $V_3=\int \frac{d^3k}{(2\pi)^3}$. 
If we consider a more general perturbation such as $\phi(t,x,y,z,u)$, $V_3$ should be replaced by $\int \frac{d^3k}{(2\pi)^3}$. 
We are now focusing on the mode $k=0$. \par
In a long wave-length limit, any field theory can be effectively described by hydrodynamics.  
We assume that the energy-momentum tensor of the boundary field theory has the following form 
\begin{align}
T^{\mu \nu}=(\epsilon+P)u^{\mu}u^{\nu}+P\eta^{\mu \nu}+\tau^{\mu \nu},
\end{align}
with energy density $\epsilon$, pressure $P$ and velocity field $u^\mu$. 
The boundary field theory is supposed to be on the four dimensional uncurved space-time. Hence, we have $\mu=0,1,2,3$ and 
$\eta^{\mu \nu}$ is Minkowski metric. The term $\tau_{\mu\nu}$ contains derivatives. Because our main interest is the first order hydrodynamic limit, 
we consider only first order derivatives. In the rest frame, $\tau_{\mu\nu}$ has no time component ($\mu=0$) and spatial components are given by 
\begin{align}
\tau_{ij}=-\eta \Big(\partial_iu_j+\partial_ju_i-\frac{2}{3}\delta_{ij}\partial_ku^k\Big)-\zeta \delta_{ij}\partial_ku^k \quad (i,j,k=1,2,3). 
\end{align}
$\eta$ and $\zeta$ represent transport coefficients called sheer viscosity and bulk vicosity.\par
Here, we assume that the fluid is firstly at rest $u^{\mu}=(1,0,0,0)$, and then the background space-time is slightly distorted $\eta_{\mu\nu}\rightarrow g_{\mu\nu}=\eta_{\mu\nu}+\delta g_{\mu\nu}$. 
Using a projection operator $\mathcal{P}^{\mu\nu}=g^{\mu \nu}+u^\mu u^\nu$, the energy-momentum tensor is written as 
\begin{align}
T^{\mu\nu}=(\epsilon+P)u^\mu u^\nu +Pg^{\mu\nu}-\mathcal{P}^{\mu \lambda}\mathcal{P}^{\nu \rho}\bigg[\eta \Big(\nabla_\lambda u_\rho+\nabla_\rho u_\lambda-\frac{2}{3}g_{\lambda \rho}\partial_\sigma u^\sigma \Big)+\zeta g_{\lambda \rho}\nabla_\sigma u^\sigma \bigg].
\end{align}
We calculate the linear response of this tensor, but our main interest is a perturbation of the type Eq.(\ref{spur}). Hence, we set 
\begin{align}
\delta g_{\mu \nu}=\left(
\begin{array}{cccc}
 0&0 &0 &0  \\
 0&0 &\delta g_{xy}(t) &0  \\
 0&\delta g_{xy}(t) &0 &0  \\
 0&0 &0 &0 
\end{array}
\right).
\end{align} 
The velocity field $u^\mu=(1,0,0,0)$ is not changed because of parity symmetry. We can easily calculate the linear level response. After the Fourier transformation, the result is
\begin{align}
\delta T^x_y=-P\delta g^x_y+i\omega \eta \delta g^x_y. \label{EMCFT}
\end{align}
We compare Eq.(\ref{EMAds}) with Eq.(\ref{EMCFT}) and  $\phi^{(0)}$ with $\delta g^x_{y}$, from which 
we interpret that the boundary field theory has the pressure $P=\frac{1}{16\pi G_5}\frac{r_0^4}{L^5}$ and the shear viscosity $\eta=\frac{1}{16\pi G_5}\big(\frac{r_0}{L}\big)^3$. \par
On the Black Hole horizon, $(x,y,z)$ components of the metric can be written as $ds^2=\big(\frac{r_0}{L}\big)^2(dx^2+dy^2+dz^2)$. 
Then, we apply the area law of Black Hole entropy. We can calculate the entropy density as $s=\frac{1}{4G_5}\big(\frac{r_0}{L}\big)^3$ which is interpreted as the entropy density of the boundary field theory. 
Thus, we obtain the ratio $\eta/s=1/4\pi$. \par
The pressure can be calculated in a different way. If we consider only background SAdS-BH with no perturbation, the Euclidean on-shell action is given by 
\begin{align}
S_E=-\frac{1}{16\pi G_5}\frac{r_0^4}{L^5}\int^\beta_0d\tau \int dxdydz=-\frac{\beta V_3}{16\pi G_5}\frac{r_0^4}{L^5}.
\end{align}
Thus, we have the partition function $Z=e^{-S_E}$ and we can calculate the pressure 
\begin{align}
P=\frac{1}{\beta}\partial_{V_3}\ln Z=\frac{1}{16\pi G_5}\frac{r_0^4}{L^5} \label{bgp},
\end{align} 
which is compatible with the value obtained from Eq.(\ref{EMAds}). \par
This is the standard calculation of the first order hydrodynamics via AdS/CFT correspondence. 
Other types of perturbations leads to other coefficients, but we will not treat them in this paper.

\subsection{the case of massive gravity}
In this subsection, we consider dRGT massive garavity and extend the prescription of the AdS/CFT correspondence in general relativity. 
We see that mass of a graviton generate extra divergences, and how to cancel them is a main topic. 
We show that not only a new counterterm have to be added but also a condition on graviton's mass must be imposed. 
For notational simplicity, we set $16\pi G_5=1$, $L=r_0=1$ and $V_3=1$ in this subsection. \par
In order to introduce mass of a graviton which is denoted as $m$, we add a interacion (mass) term $S_{int}$ to the action of general relativity. 
Then, the action of dRGT massive gravity is given by 
\begin{align}
&S=S_{EH}+S_{GH}+S_{ct}+S_{int}\\
&S_{int}=m^2\int d^5x \sqrt{-g}\, e\big(\sqrt{g^{-1}\bar{g}}\big), \label{mSint}
\end{align}
where $\bar{g}$ is a background metric and $g$ is a full metric (background +fluctuation) $g=\bar{g}+\delta g$. In this paper, we use SAdS-BH as a background metric $\bar{g}=$(SAdS-BH). 
Einstein-Hilbert action $S_{EH}$, Gibbons-Hawking term $S_{GH}$ and the counterterm $S_{ct}$ are the same as those of general relativity. They are all constructed from the full metric $g$. 
The difference comes from the term $S_{int}$ which depends explicitly on $\bar{g}$. This background dependence breaks the diffeomorphism invariance. 
The mass term $S_{int}$ is composed of the function 
$e\big(\sqrt{g^{-1}\bar{g}}\big)$. This is a function of a matrix ${(\sqrt{g^{-1}\bar{g}})^{\mu}}_{\nu}$ where the square root of a matrix means 
${(\sqrt{g^{-1}\bar{g}})^{\mu}}_{\lambda}{(\sqrt{g^{-1}\bar{g}})^{\lambda}}_{\nu}={(g^{-1}\bar{g})^{\mu}}_{\nu}=g^{\mu \lambda}\bar{g}_{\lambda \nu}$. 
The explicit formula of this function is \cite{Hassan:2011vm} \cite{Hassan:2011tf} 
\begin{align}
e(A)=\sum_{n=0}^{5}\beta_n\epsilon_{\mu_1\cdots \mu_n \lambda_{n+1} \cdots \lambda_5}\epsilon^{\nu_1\cdots \nu_n \lambda_{n+1} \cdots \lambda_5}A^{\mu_1}_{\nu_1}\cdots A^{\mu_n}_{\nu_n},
\end{align}
where $\epsilon$ is an antisymmetric tensor. The function $e$ contains constants $\beta_n$ which are adjusted to satisfy the relation 
$e(1)=0$ (1 is a unit matrix) and reduce to the Pauli-Fiertz mass term in the expansion up to the second order in $\delta g$
\begin{align}
S_{int}&=-\frac{1}{4}m^2\int d^5x \sqrt{-\bar{g}}\Big(\text{Tr}(\delta g)^2-\text{Tr}^2(\delta g)\Big),
\end{align}
where we abbreviated as $\text{Tr}^2A=\text{Tr}A \times \text{Tr}A$ and $\text{Tr}A^2$ is a tarce of the matrix $A^2$.\par
Now, we take a perturbation of the same type as Eq.(\ref{spur}) and expand the action up to the second order in $\phi$. 
The calulation is almost the same as that of general relativity in the previous subsection. 
The only difference is that $S_{int}$ is added to the bulk action Eq.(\ref{Sbulk})
\begin{align}
S_{bulk}=S_{EH}+S_{int}=\int du \, \mathcal{L}(\phi \, ,\, \phi' \, ,\, \phi''),
\end{align}
where
\begin{align}
S_{int}&=-\int \frac{d\omega}{2\pi}\int^1_0 du \frac{m^2}{2}\frac{1}{u^5}\phi_{-\omega}\phi_\omega.
\end{align}
Because $S_{int}$ contains no derivative, Eq.(\ref{Sbulkbd}) is not changed and the diffenrence occurs only in the equation of motion Eq.(\ref{EOM}). 
\begin{align}
\text{EOM : }& \Big(\frac{h}{u^3}\phi'_\omega \Big)'-m^2\frac{1}{u^5}\phi_{\omega}+\frac{\omega^2}{u^3h}\phi_\omega=0. \label{mEOM}
\end{align}
There is no change in Eq.(\ref{Sren})
\begin{align}
S=S_{bulk}+S_{GH}+S_{ct}&=V_4+\int \frac{d\omega}{2\pi}\bigg\{ \Big(-\frac{1}{2}+O[u^4]\Big)\phi_{-\omega} \phi_{\omega} +\frac{1}{2}\Big(\frac{1}{u^3}-u\Big)\phi_{-\omega}\phi'_{\omega}\bigg\} \Bigg|_{u=0}. \label{mS}
\end{align}
We solve the EOM Eq.(\ref{mEOM}) up to the first order expansion in $\omega$ 
\begin{align}
\phi_{\omega}(u)=A_{\omega}\Big\{u^{2-2\alpha}+\frac{1}{4}(1-\alpha)u^{6-2\alpha}+O[u^{10-2\alpha}]\Big\} +B_\omega \Big\{u^{2+2\alpha}+\frac{1}{4}(1+\alpha)u^{6+2\alpha}+O[u^{10+2\alpha}]\Big\}, \label{ms}
\end{align}
where we set $A_{\omega}=A_0+\omega A_1$ and $B_\omega=B_0+\omega B_1$. $A_{0,1}$ and $B_{0,1}$ are constants as the previous subsection. Mass of a graviton is contained in $\alpha=\sqrt{1+m^2/4}$. 
We substitute the solution of the EOM Eq.(\ref{ms}) into the action Eq.(\ref{mS}) and obtain 
\begin{align}
S=V_4+\int \frac{d\omega}{2\pi}\bigg\{&(1+\alpha)A_{-\omega}B_{\omega}+(1-\alpha)B_{-\omega}A_{\omega} \\
                                      &+A_{-\omega}A_{\omega}\Big((1-\alpha)u^{-4\alpha}+\frac{1}{2}(\alpha^2-\alpha-1)u^{4-4\alpha}+O[u^{8-4\alpha}]\Big) \bigg\} \Bigg|_{u=0}. 
\end{align}
Hence, we see that extra divergences arise from $u^{-4\alpha}|_{u=0}, u^{4-4\alpha}|_{u=0}$ and $O[u^{8-4\alpha}]|_{u=0}$. \par
Here, it should be noted that if we consider pure AdS space-time, we set $h=1$ in the metric of SAdS-BH Eq.(\ref{metAdSBH}). 
Setting $h=1$ in the EOM Eq.(\ref{mEOM}), the solution becomes $\phi_{\omega}(u)=A_{\omega}u^{2-2\alpha}+B_\omega u^{2+2\alpha}$. 
Higher order terms such as $O[u^{6-2\alpha}]$ or $O[u^{6+2\alpha}]$ in Eq.(\ref{ms}) come from the expansion of $h=1-u^4$ around $u\sim 0$. 
In addition, if we set $h=1$ in Eq.(\ref{hSbulk}), Eq.(\ref{hSGH}) and Eq.(\ref{hSct}), the action Eq.(\ref{mS}) becomes 
$S =V_4+\int \frac{d\omega}{2\pi} \frac{1}{2u^3}\phi_{-\omega}\phi_{\omega}'|_{u=0}$. 
Therfore, on pure AdS space-time, only the divergence $u^{-4\alpha}|_{u=0}$ occurs. \par
In order to cancel these divergences, we attempt to add a new counterterm $S_{mct}$. 
This situation resembles the case of a massive scalar field \cite{Skenderis:2002wp}, where extra 
divergences can be canceled by a mass term on the AdS-boundary. Therefore, we try to add a term 
\begin{align}
S_{mct} \propto \int_{Ads-bdy} d^4 x \sqrt{-\gamma} \, e\big(\gamma^{-1}\bar{\gamma}\big), \label{Smct}
\end{align}
which reduces to the Pauli-Fiertz mass term in the second order expansion 
\begin{align}
S_{mct}=-\frac{1}{2}(1-\alpha)\int_{Ads-bdy} d^4 x \sqrt{-\bar{\gamma}}\Big(\text{Tr}(\delta \gamma)^2-\text{Tr}^2(\delta \gamma)\Big). \label{SmctPF}
\end{align}
The coefficient $(1-\alpha)$ is adjusted for our purpose. 
If we consider a perturbation dependent on spatial coodinates $(x,y,z)$, we may need other counterterms \cite{Skenderis:2002wp}. However, we do not treat this topic in this paper. 
In appendix, we investigate the validity of this counterterm in a different perturbation. \par 
Inserting the perturbation Eq.(\ref{spur}) and the solution of the EOM Eq.(\ref{ms}), we have 
\begin{align}
S_{mct}=&-\big(1-\alpha \big)\int_{Ads-bdy} d^4 x \sqrt{-\gamma}\phi^2\\
      =&-\big(1-\alpha \big)\int \frac{d\omega}{2\pi}\frac{\sqrt{1-u^4}}{u^4}\phi_{-\omega}\phi_{\omega} \Big|_{u=0}\\
      =&-\big(1-\alpha \big)\int \frac{d\omega}{2\pi}\Big(\frac{1}{u^4}-\frac{1}{2}+O[u^4]\Big)\phi_{-\omega}\phi_{\omega} \Big|_{u=0}\\
      =&\int \frac{d\omega}{2\pi}\bigg\{(\alpha-1)(A_{-\omega}B_{\omega}+B_{-\omega}A_{\omega})\\
       & \qquad \qquad +A_{-\omega}A_{\omega}\Big(-(1-\alpha)u^{-4\alpha}+\frac{1}{2}\alpha(1-\alpha)u^{4-4\alpha}+O[u^{8-2\alpha}]\Big)\bigg\}\Bigg|_{u=0}.
\end{align}
Thus, we obtain 
\begin{align}
S+S_{mct}=V_4+\int \frac{d\omega}{2\pi}\Big\{2\alpha A_{-\omega}B_{\omega}+A_{-\omega}A_{\omega}\Big(-\frac{1}{2}u^{4-4\alpha}+O[u^{8-2\alpha}]\Big)\Big\}\Big|_{u=0}. \label{SSmct}
\end{align}
The divergence from $u^{-4\alpha}$ has been canceled, but still other divergences remain. 
If we take pure AdS space-time as a background, these remaining divergences do not appear and $S_{mct}$ is enough. 
Cancelation of them requires a condition on graviton's mass. 
We have to set $-4<m^2<0$, namely $0<\alpha<1$, which is a reminiscence of the BF-bound \cite{Breitenlohner:1982bm, Breitenlohner:1982jf, Mezincescu:1984ev}. 
BF-bound is a result of the stability analysis, so we should investigate the stability in massive gravity. However, we left this issue for a future work and continue our calculation. \par
Then, the non-divergent on-shell action in dRGT massive gravity is given by
\begin{align}
S+S_{mct}=V_4+\int \frac{d\omega}{2\pi}(2\alpha A_{-\omega}B_{\omega}). 
\end{align} \par
We attempt to fix constants $A_{0,1}$ and $B_{0,1}$. If we assume that the massive and massles solution Eq.(\ref{ms}) and Eq.(\ref{np}) coinside in the massless limit $\alpha=1$, 
we should set $A_{\omega}=\phi_\omega^{(0)}$ and $B_{\omega}=\frac{i\omega}{4}\phi_\omega^{(0)}$, which leads to 
\begin{align}
&S+S_{mct}=V_4+\int \frac{d\omega}{2\pi}\Big(\frac{i\alpha \omega}{2}\Big)\phi_{-\omega}^{(0)}\phi_\omega^{(0)}\\
&<\delta T^{xy}_{\omega}>=\frac{\delta S}{\delta \phi^{(0)}_{-\omega}}=i\omega \alpha \phi^{(0)}_{\omega}.
\end{align}
Compared to Eq.(\ref{EMCFT}), $P=0$ and $\eta=\alpha$. The pressure is zero which is not consistent with the value calculated from the background metric Eq.(\ref{bgp}). 
We do not know how to interpret this result. 
We suspect that this peculiarity comes from the weird feature of dRGT massive gravity, where the mass term depends explicitly on the back groundmetric. 
This feature breaks the diffeomorphism invariance. Therfore, it seems natural to promote this background metric to another dynamical variable, which is nothing but bimetric gravity. 
We expect that the peculiarity in dRGT massive gravity is modified in bimetric gravity. We consider this topic in the next section.

\section{bimetric gravity and the AdS/CFT correspondence in the first order hydrodynamic limit}\label{bimetricgravity}
In this section, we extend the method of the previous section to the case of bimetric gravity. Our interest is what emerges on the boundary field theory. 
In section \ref{massivegravity}, we treated dRGT massive gravity, but it has one peculiar feature. 
It explicitly contains a reference (background) metric, which breaks the diffeomorphism invariance. Hence, it seems natural to make 
the reference metric dynamical and revive the invariance. This is nothing but bimetric gravity \cite{Hassan:2011ea, Hassan:2011zd}. 
Therefore, bimetric gravity is a generalization of dRGT massive gravity, which contains two metrics. 
In the following, we write these metrics as $g$ and $f$, and their induced metrics on the AdS-boundary as $\gamma$ and $\rho$ respectively. \par
Then, we start with the action given by 
\begin{align}
S=&S_{EH}[g]+S_{GH}[\gamma]+S_{ct}[\gamma]\\ 
 +&S_{EH}[f]+S_{GH}[\rho]+S_{ct}[\rho]\\
 +&S_{int}[g,f]+S_{int,ct}[\gamma,\rho], 
\end{align}
where 
\begin{align}
&S_{EH}[g]+S_{GH}[\gamma]+S_{ct}[\gamma]\nonumber \\
&=M_g^2\int d^5x \sqrt{-g} \big(R[g]-2\Lambda)+2M_g^2\int_{AdS-bdy} d^4x \sqrt{-\gamma}K[\gamma]+M_g^2\int_{AdS-bdy} d^4x \sqrt{-\gamma}\Big(\frac{6}{L}+\cdots \Big)
\end{align}
and 
\begin{align}
&S_{EH}[f]+S_{GH}[\rho]+S_{ct}[\rho]\nonumber \\
&=M_f^2\int d^5x \sqrt{-f} \big(R[f]-2\Lambda)+2M_f^2\int_{AdS-bdy} d^4x \sqrt{-\rho}K[\rho]+M_f^2\int_{AdS-bdy} d^4x \sqrt{-\rho}\Big(\frac{6}{L}+\cdots \Big).
\end{align}
In bimetric gravity, we can introduce different Planck masses (Gravitational constants) for the two metrics, which we write as $M_g$ and $M_f$. 
$R[g]$ is the scalar curvature for $g_{\mu \nu}$ and $R[f]$ is the scalar curvature for $f_{\mu \nu}$. $K[\gamma]$ and $K[\rho]$ represent the extrinsic curvatures for each metric. 
In general, cosmological constants for $g_{\mu \nu}$ and $f_{\mu \nu}$ can be different. 
However, in this paper, we assume that they have the same value $\Lambda$ and each mertric has the same AdS-radius $L$. 
We impose this condition in order to take perturbations on the same background (SAdS-BH). 
The interaction term $S_{int}[g,f]$ is a genaralization of $S_{int}$ in dRGT massive gravity Eq.(\ref{mSint}) 
\begin{align}
S_{int}[g,f]=2m^2M_{eff}^2\int d^5x \sqrt{-g}\, e\big(\sqrt{g^{-1}f}\big),
\end{align}
where background metric $\bar{g}$ in Eq.(\ref{mSint}) is raplaced by the other dynamical metric $f$, and $M_{eff}$ is defined as 
\begin{align}
M_{eff}^2=\Big(\frac{1}{M_g^2}+\frac{1}{M_f^2}\Big)^{-1}.
\end{align}
The counterterm $S_{int,ct}[\gamma, \rho]$ is a extension of the counterterm we added in maissive gravity Eq.(\ref{Smct}) 
\begin{align}
S_{int,ct}[\gamma,\rho] \propto \frac{M_{eff}^2}{L}\int_{AdS-bdy} d^4 x \sqrt{-\gamma} \, e\big(\sqrt{\gamma^{-1}\rho}\big).
\end{align}
Because the function $e$ satisfies $e(1)=0$, $S_{int}[g,f]$ vanishes when we set $f=g$. Thus, we have a solution $g=f=$(SAdS-BH). 
In the following, we consider a perturbation $g=\bar{g}+\delta g$ and $f=\bar{f}+\delta f$ on the same background $\bar{g}=\bar{f}=$(SAdS-BH). 
Thus, the expansion of $S_{int}[g,f]$ and $S_{int,ct}[\gamma,\rho]$ up to the second order in $\delta g$ and $\delta f$ is given by  
\begin{align}
S_{int}[g,f]&=-\frac{1}{4}m^2M_{eff}^2\int d^5x \sqrt{-\bar{g}}\Big(\text{Tr}(\delta g-\delta f)^2-\text{Tr}^2(\delta g-\delta f)\Big)\\
S_{int,ct}[\gamma,\rho]&=-\frac{1}{2}(1-\alpha)\frac{M_{eff}^2}{L}\int_{Ads-bdy} d^4 x \sqrt{-\bar{\gamma}}\Big(\text{Tr}(\delta \gamma-\delta \rho)^2-\text{Tr}^2(\delta \gamma-\delta \rho)\Big),
\end{align}
where we put $\alpha=\sqrt{1+(mL)^2/4}$. The coefficient $(1-\alpha)$ is adjusted to cancel the divergence coming from $u^{-4\alpha}|_{u=0}$. \par
Now, we take a perturbation such as Eq.(\ref{spur}) 
\begin{align}
&{\delta g^{\mu}}_{\nu}=\bar{g}^{\mu \lambda}\delta g_{\lambda \nu}=\left(
\begin{array}{ccccc}
0 & 0& 0& 0& 0 \\
0 & 0& \phi&0 & 0 \\
0 & \phi& 0& 0&0  \\
0 & 0& 0& 0& 0 \\
0 & 0& 0& 0& 0
\end{array}
\right)  ,\qquad \phi=\phi(t,u)\\
&{\delta f^{\mu}}_{\nu}=\bar{g}^{\mu \lambda}\delta f_{\lambda \nu}=\left(
\begin{array}{ccccc}
0 & 0& 0& 0& 0 \\
0 & 0& \psi&0 & 0 \\
0 & \psi& 0& 0&0  \\
0 & 0& 0& 0& 0 \\
0 & 0& 0& 0& 0
\end{array}
\right)  ,\qquad \psi=\psi(t,u).
\end{align} 
To begin with, we calculate $S_{int}$ and $S_{int,ct}$ 
\begin{align}
S_{int}&=-\frac{r_0^4}{L^5}M_{eff}^2\frac{(mL)^2}{2}V_3\int \frac{d\omega}{2\pi}\int^1_0 du \frac{1}{u^5}(\phi_{-\omega}-\psi_{-\omega})(\phi_\omega-\psi_\omega) \\
S_{int,ct}&=-(1-\alpha)\frac{r_0^4}{L^5}M_{eff}^2V_3 \int \frac{d\omega}{2\pi} \frac{\sqrt{h}}{u^4}(\phi_{-\omega}-\psi_{-\omega})(\phi_\omega-\psi_\omega)\Big|_{u=0}.
\end{align}
The bulk action $S_{bulk}=S_{EH}[g]+S_{EH}[f]+S_{int}[g,f]$ is expanded as  
\begin{align}
S_{bulk}=&-\frac{r_0^4}{L^5}(M_g^2+M_f^2)V_4\int^1_0 du \frac{8}{u^5} \nonumber \\
 +&\frac{r_0^4}{L^5}M_g^2V_3\int \frac{d\omega}{2\pi}\int^1_0 du \bigg\{\frac{3}{2}\frac{h}{u^3}\phi'_{-\omega} \phi'_{\omega}+2\frac{h}{u^3}\phi_{-\omega} \phi''_\omega-\frac{8}{u^4}\phi_{-\omega} \phi'_{\omega}+\bigg(\frac{1}{2u^3h}\Big(\frac{L^2}{r_0}\omega \Big)^2+\frac{4}{u^5}\bigg)\phi_{-\omega}\phi_{\omega}\bigg\} \nonumber \\
 +&\frac{r_0^4}{L^5}M_f^2V_3\int \frac{d\omega}{2\pi}\int^1_0 du \bigg\{\frac{3}{2}\frac{h}{u^3}\psi'_{-\omega} \psi'_{\omega}+2\frac{h}{u^3}\psi_{-\omega} \psi''_\omega-\frac{8}{u^4}\psi_{-\omega} \psi'_{\omega}+\bigg(\frac{1}{2u^3h}\Big(\frac{L^2}{r_0}\omega \Big)^2+\frac{4}{u^5}\bigg)\psi_{-\omega}\psi_{\omega}\bigg\}\nonumber \\
 -&\frac{r_0^4}{L^5}M_{eff}^2\frac{(mL)^2}{2}V_3\int \frac{d\omega}{2\pi}\int^1_0 du \frac{1}{u^5}(\phi_{-\omega}-\psi_{-\omega})(\phi_\omega-\psi_\omega).
\end{align}
Because interaction term $S_{int}$ contains no derivative, we have 
\begin{align}
S=& \frac{r_0^4}{L^5}V_4(M_g^2+M_f^2)\\
  +& \frac{r_0^4}{L^5}M_g^2V_3 \int \frac{d\omega}{2\pi}\bigg\{ \Big(-\frac{1}{2}+O[u^4]\Big)\phi_{-\omega} \phi_{\omega} +\frac{1}{2}\Big(\frac{1}{u^3}-u\Big)\phi_{-\omega}\phi'_{\omega}\bigg\} \Bigg|_{u=0}\\
      +& \frac{r_0^4}{L^5}M_f^2V_3\int \frac{d\omega}{2\pi}\bigg\{ \Big(-\frac{1}{2}+O[u^4]\Big)\psi_{-\omega} \psi_{\omega} +\frac{1}{2}\Big(\frac{1}{u^3}-u\Big)\psi_{-\omega}\psi'_{\omega}\bigg\} \Bigg|_{u=0}\\
      -& \big(1-\alpha \big)\frac{r_0^4}{L^5}M_{eff}^2V_3 \int \frac{d\omega}{2\pi}\Big(\frac{1}{u^4}-\frac{1}{2}+O[u^4]\Big)(\phi_{-\omega}-\psi_{-\omega})(\phi_{\omega}-\psi_{\omega}) \Big|_{u=0}.
\end{align}
Here, we put $\tilde{\phi}=Mg\phi$ and $\tilde{\psi}=M_f\psi$, and we also define 
\begin{align}
\frac{1}{M_{eff}}\Phi=\frac{\tilde{\phi}}{M_f}+\frac{\tilde{\psi}}{M_g}\, , \qquad  \frac{1}{M_{eff}}\Psi=\frac{\tilde{\phi}}{M_g}-\frac{\tilde{\psi}}{M_f}.
\end{align}
Using the relattion such as $\tilde{\phi}^2+\tilde{\psi}^2=\Phi^2+\Psi^2$, we obtain 
\begin{align}
S=& \frac{r_0^4}{L^5}V_4(M_g^2+M_f^2)\\
 +& \frac{r_0^4}{L^5}V_3\int \frac{d\omega}{2\pi}\bigg\{ \Big(-\frac{1}{2}+O[u^4]\Big)\Phi_{-\omega} \Phi_{\omega} +\frac{1}{2}\Big(\frac{1}{u^3}-u\Big)\Phi_{-\omega}\Phi'_{\omega}\bigg\} \Bigg|_{u=0}\\
 +& \frac{r_0^4}{L^5}V_3\int \frac{d\omega}{2\pi}\bigg\{ \Big(-\frac{1}{2}+O[u^4]\Big)\Psi_{-\omega} \Psi_{\omega} +\frac{1}{2}\Big(\frac{1}{u^3}-u\Big)\Psi_{-\omega}\Psi'_{\omega}\bigg\} \Bigg|_{u=0}\\
 -& \big(1-\alpha \big)\frac{r_0^4}{L^5}V_3\int \frac{d\omega}{2\pi}\Big(\frac{1}{u^4}-\frac{1}{2}+O[u^4]\Big)\Psi_{-\omega}\Psi_{\omega} \Big|_{u=0}.
\end{align}
We also obtain the EOM for $\Phi$ and $\Psi$ from the bulk action $S_{bulk}$
\begin{align}
\text{EOM : }& \Big(\frac{h}{u^3}\Phi'_\omega \Big)'+\Big(\frac{L^2}{r_0}\omega^2\Big)^2\frac{1}{u^3h}\Phi_\omega=0\\
             & \Big(\frac{h}{u^3}\Psi'_\omega \Big)'-(Lm)^2\frac{1}{u^5}\Psi_{\omega}+\Big(\frac{L^2}{r_0}\omega^2\Big)^2\frac{1}{u^3h}\Psi_\omega=0.
\end{align}
We solve the EOM up to the first order expansion in $\omega$
\begin {align}
\Phi_\omega(u)&= A_{\omega}+B_{\omega}(u^4+O[u^8])\\
\Psi_{\omega}(u)&=C_{\omega}\Big\{u^{2-2\alpha}+\frac{1}{4}(1-\alpha)u^{6-2\alpha}+O[u^{10-2\alpha}]\Big\} +D_\omega \Big\{u^{2+2\alpha}+\frac{1}{4}(1+\alpha)u^{6+2\alpha}+O[u^{10+2\alpha}]\Big\}, 
\end{align}
where $A_\omega$, $B_\omega$, $C_\omega$ and $D_\omega$ are constants and can be written as $A_\omega=A_0+\omega A_1$, $B_\omega=B_0+\omega B_1$, etc.
Then, we obtain the on-shell action 
\begin{align}
S=\frac{r_0^4}{L^5}V_4(M_g^2+M_f^2)+\frac{r_0^4}{L^5}V_3\int \frac{d\omega}{2\pi}\Big(-\frac{1}{2}A_{-\omega}A_\omega +2A_{-\omega}B_\omega \Big) +\frac{r_0^4}{L^5}V_3\int \frac{d\omega}{2\pi}(2\alpha C_{-\omega}D_{\omega}), \label{BiS}
\end{align}
where we imposed a condition $0<\alpha<1$ to make the terms $O[u^{4-4\alpha}]|_{u=0}$ finite. \par
We have obtained the on-shell action Eq.(\ref{BiS}) written by massless and massive modes $\Phi$ and $\Psi$. 
However, they are mixture of original variables $g_{\mu \nu}$ and $f_{\mu \nu}$, and we do not know how to interpret this mixture in the context of gravity/fluid correspondence. 
Hence, we proceed our calculation with the variables $g_{\mu \nu}$ and $f_{\mu \nu}$ (or $\phi$ and $\psi$). 
In addition, we have to consider a boundary condition on the Black Hole horizon to fix the constants $A$, $B$, $C$ and $D$. 
In the massless limit $m=0$, bimetric gravity decouoles to a pair of independent genaral relativities, in which case we should select the ingoing wave condition. 
Therefore, it seems natural for the solutions of $\phi$ and $\psi$ to mache Eq.(\ref{np}) in the massless limit. \par
\begin{figure}[t]
\begin{minipage}{0.4\textwidth}
\begin{center}
\includegraphics[scale=0.25]{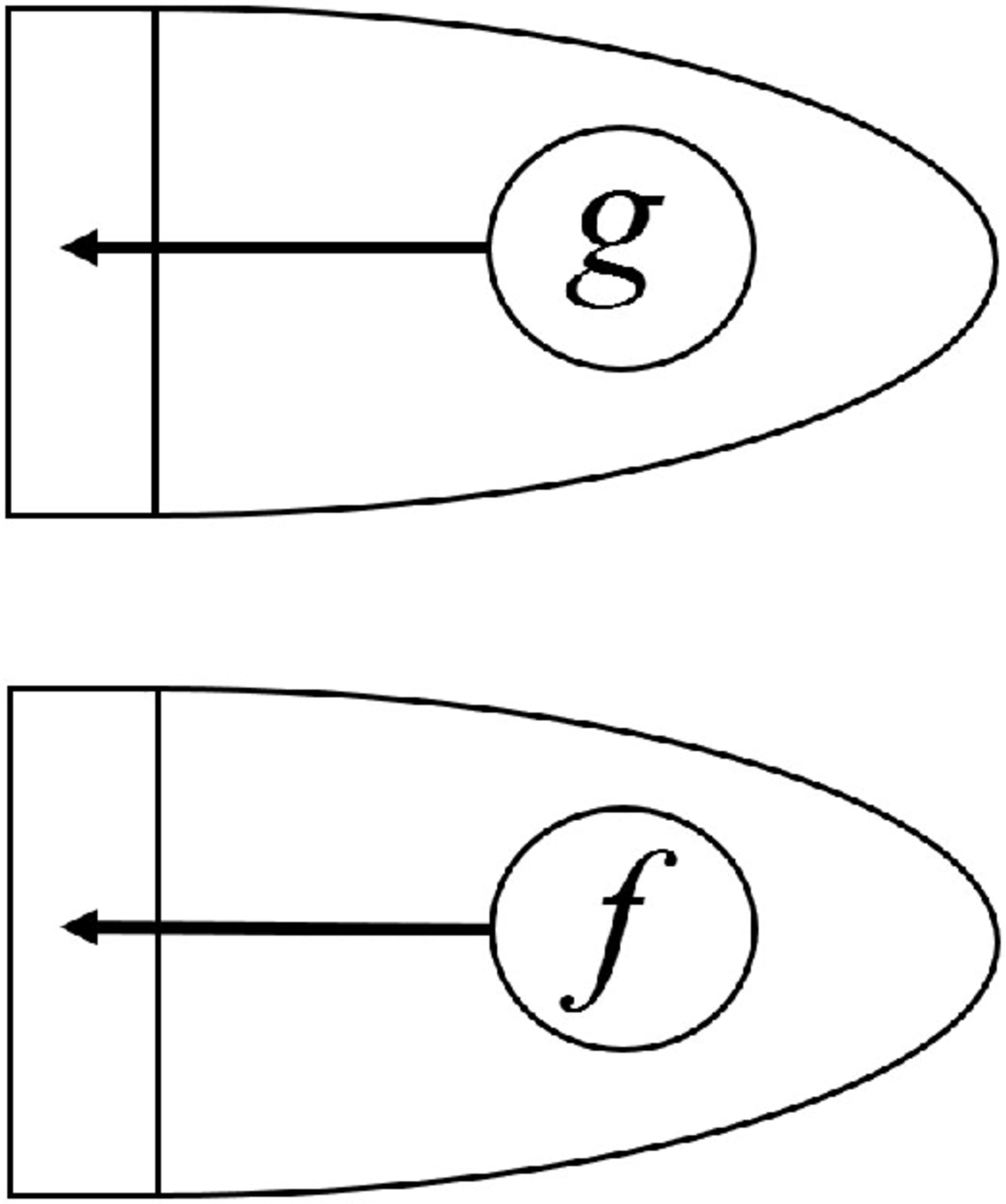}
\end{center}
\caption{non-interacting case}
\label{nonint}
\end{minipage}
\begin{minipage}{0.4\textwidth}
\begin{center}
\includegraphics[scale=0.25]{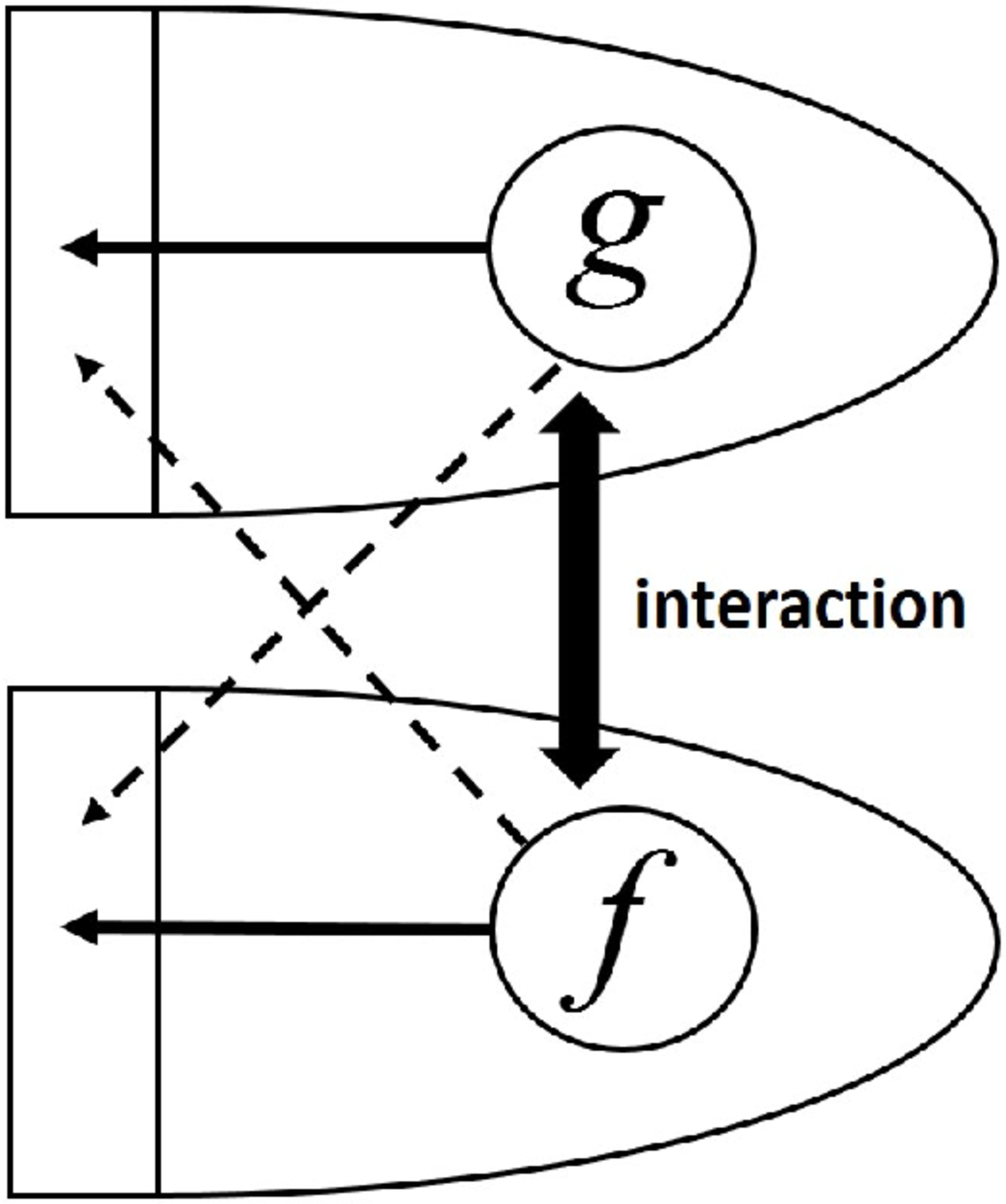}
\end{center}
\caption{interacting case}
\label{int}
\end{minipage}
\end{figure}

If we set $\alpha=1$ ($m^2=0$), we have 
\begin{align}
\phi_{\omega}=\frac{M_g \Phi_{\omega}+M_f \Psi_{\omega}}{M_g\sqrt{M_g^2+M_f^2}}=\frac{M_g A_{\omega}+M_f C_{\omega}}{M_g\sqrt{M_g^2+M_f^2}}+\frac{M_g B_{\omega}+M_f D_{\omega}}{M_g\sqrt{M_g^2+M_f^2}}u^4+O[u^8]\\
\psi_{\omega}=\frac{M_f \Phi_{\omega}-M_g\Psi_{\omega}}{M_f\sqrt{M_g^2+M_f^2}}=\frac{M_f A_{\omega}-M_g C_{\omega}}{M_f\sqrt{M_g^2+M_f^2}}+\frac{M_f B_{\omega}-M_g D_{\omega}}{M_f\sqrt{M_g^2+M_f^2}}u^4+O[u^8].
\end{align}
Thus, we put
\begin{align}
& \phi_\omega ^{(0)}=\frac{M_g A_{\omega}+M_f C_{\omega}}{M_g\sqrt{M_g^2+M_f^2}}\, ,\qquad i\frac{L^2\omega}{4r_0}\phi_{\omega}^{(0)}=\frac{M_g B_{\omega}+M_f D_{\omega}}{M_g\sqrt{M_g^2+M_f^2}}\\
& \psi_\omega ^{(0)}=\frac{M_f A_{\omega}-M_g C_{\omega}}{M_f\sqrt{M_g^2+M_f^2}}\, ,\qquad i\frac{L^2\omega}{4r_0}\psi_{\omega}^{(0)}=\frac{M_f B_{\omega}-M_g D_{\omega}}{M_f\sqrt{M_g^2+M_f^2}}, 
\end{align}
and we obtain 
\begin{align}
&A_{\omega}=\frac{M_g^2\phi_\omega^{(0)}+M_f^2\psi_\omega^{(0)}}{\sqrt{M_g^2+M_f^2}}\, ,\qquad B_\omega=\Big(i\frac{L^2\omega}{4r_0}\Big)\frac{M_g^2\phi_\omega^{(0)}+M_f^2\psi_\omega^{(0)}}{\sqrt{M_g^2+M_f^2}}\\
&C_\omega=\frac{M_gM_f(\phi_\omega^{(0)}-\psi_\omega^{(0)})}{\sqrt{M_g^2+M_f^2}}\, ,\qquad D_\omega=\Big(i\frac{L^2\omega}{4r_0}\Big)\frac{M_gM_f(\phi_\omega^{(0)}-\psi_\omega^{(0)})}{\sqrt{M_g^2+M_f^2}}.
\end{align}
Inserting these relations into Eq.(\ref{BiS}), we obtain the on-shell action written by original variables $g_{\mu \nu}$ and $f_{\mu \nu}$ ($\phi$ and $\psi$) 
\begin{align}
S=\frac{r_0^4}{L^5}V_4(M_g^2+M_f^2)\qquad \qquad & \\
+\frac{r_0^4}{L^5}\Big(\frac{1}{M_g^2+M_f^2}\Big)V_3\int \frac{d\omega}{2\pi}\bigg\{&-\frac{1}{2}M_g^4\phi^{(0)}_{-\omega}\phi_\omega^{(0)} +i\frac{L^2\omega}{2r_0}M_g^2(M_g^2+\alpha M_f^2)\phi^{(0)}_{-\omega}\phi_\omega^{(0)}\\
                                                                                    &-\frac{1}{2}M_f^4\psi^{(0)}_{-\omega}\psi_\omega^{(0)} +i\frac{L^2\omega}{2r_0}M_f^2(M_f^2+\alpha M_g^2)\psi^{(0)}_{-\omega}\psi_\omega^{(0)}\\
                                       &-\frac{1}{2}M_g^2M_f^2\Big(\phi^{(0)}_{-\omega}\psi_\omega^{(0)}+\psi^{(0)}_{-\omega}\phi_\omega^{(0)}\Big)\\
                                       &+i\frac{L^2 \omega}{2r_0}M_g^2M_f^2(1-\alpha)\Big(\phi^{(0)}_{-\omega}\psi_\omega^{(0)}+\psi^{(0)}_{-\omega}\phi_\omega^{(0)}\Big)\bigg\}. 
\end{align}
This on-shell action contains mixed terms such as $\phi\psi$, which suggests the emergence of two-component fluid. 
If the metrics $g$ and $f$ do not interact, we have two independent AdS (bulk)/CFT (boundary) pairs. The fluctuation of $g$ enters into one boundary and becomes a source to generate one field. 
The fluctuation of $f$ goes into the other boundary and becomes a source of the other field (FIG.\ref{nonint}). 
We call these boundaries as g-boundary and f-boundary for convenience. 
However, if their interaction is switched on, perturbations begin to go into not only the original boundary but also the other. For example, perturbed metric $g$ enters into f-boundary as well as g-boundary. 
As a result, two fields are generated on each boundary (FIG.\ref{int}).
\par

Now, the GKP-Witten relation can be written as 
\begin{align}
\Big \langle  \exp \Big(i \int \phi^{(0)}\mathcal{O}_g+\phi^{(0)}\mathcal{O}_f+\psi^{(0)}\mathcal{Q}_g+\psi^{(0)}\mathcal{Q}_f\Big)   \Big \rangle =\exp \Big(iS\big[\phi,\psi|_{u=0}=\phi^{(0)},\psi^{(0)}\big]\Big), \label{GKPW2}
\end{align}
where $\mathcal{O}_g$ and $\mathcal{Q}_g$ are operators on g-boundary, and $\mathcal{O}_f$ and $\mathcal{Q}_f$ are on f-boundary. 
$\phi^{(0)}$ becomes a source of not only $\mathcal{O}_g$ on g-boundary but also $\mathcal{O}_f$ on f-boundary. $\psi^{(0)}$ becomes a souce of $\mathcal{Q}_g$ as well as $\mathcal{Q}_f$. 
In our setting, these operators are interpreted as energy momentum tensors. \par
Here, we remember the discussion of general relativity. 
We considered a perturbation around SAdS-BH and obtained the expectation value of the perturbed energy momentum tensor Eq.(\ref{EMAds}) via the AdS/CFT correspondence. 
On the other hand, we focused on the boundary field theory. We assumed that the boundary space-time was slightly distorted and calculated the linear response of the energy momentum tensor Eq.(\ref{EMCFT}). 
We compared Eq.(\ref{EMAds}) with Eq.(\ref{EMCFT}), and read the coefficients. \par
Now, we proceed in the same way. 
We have two boundaries, namely g-boundary and f-boundary. Each boundary field theory has some energy momentum tensor though we do not know their concrete formulae. 
Schematically, we write them as T[g] and T[f]. T[g] is an energy momentum tensor on g-boundary and T[f] is on f-boundary. 
Then, we assume that the boundary space-times are slightly distorted and consider the linear response of these energy momentum tensors. 
If we write the distortion as $\eta_{\mu \nu}\rightarrow \eta_{\mu \nu}+\delta g_{\mu \nu}$ on g-boundary and $\eta_{\mu \nu}\rightarrow \eta_{\mu \nu}+\delta f_{\mu \nu}$ on f-boundary 
where $\eta_{\mu \nu}$ $(\mu, \nu=0,1,2,3)$ is four dimensional Minkowski metric, 
it seems natural to think that the response on g-boundary $\delta T[g]$ should consists of only $\delta g_{\mu \nu}$ and the response on f-boundary $\delta T[f]$ should consists of only $\delta f_{\mu \nu}$. 
We seek the concrete forms of these linear responses $\delta T[g] \propto \delta g$ and $\delta T[f] \propto \delta f$ from the expectation values calculated via AdS/CFT correspondence $<\mathcal{O}_{g,f}>$ and $<\mathcal{Q}_{g,f}>$. 
Thus, the expectation values on g-boundary $<\mathcal{O}_g>$ and $<\mathcal{Q}_g>$ should contain only the metric $g$ (or $\phi)$. 
The values $<\mathcal{O}_f>$ and $<\mathcal{Q}_f>$ should contain only the metric $f$ (or $\psi$). \par
Focusing on g-boundary, the expectation values of the energy momentum tensors are calculated as 
\begin{align}
& <\mathcal{O}_g>=\frac{\delta S}{\delta \phi^{(0)}_{-\omega}}\bigg|_{\psi=0}=-\Big(\frac{r_0^4}{L^5}\Big)\frac{M_g^4}{M_g^2+M_f^2}\phi_\omega^{(0)}+i\omega \Big(\frac{r_0^3}{L^3}\Big)\frac{M_g^2(M_g^2+\alpha M_f^2)}{M_g^2+M_f^2}\phi_\omega^{(0)} \label{EMphi}\\
& <\mathcal{Q}_g>=\frac{\delta S}{\delta \psi^{(0)}_{-\omega}}\bigg|_{\psi=0}=-\Big(\frac{r_0^4}{L^5}\Big)\frac{M_g^2M_f^2}{M_g^2+M_f^2}\phi_\omega^{(0)}+i\omega \Big(\frac{r_0^3}{L^3}\Big)\frac{M_g^2M_f^2(1-\alpha)}{M_g^2+M_f^2}\phi_\omega^{(0)}\label{EMpsi}.
\end{align}
These formulae and Eq.(\ref{EMCFT}) have the same form. 
Thus, we compare them and coclude that we have two-component fluid. The pressure $P$ and the sheer viscosity $\eta$ of each cmponent are given by 
\begin{align}
&P[g]_{\phi}=\Big(\frac{r_0^4}{L^5}\Big)\frac{M_g^4}{M_g^2+M_f^2}\,  , \qquad  P[g]_\psi=\Big(\frac{r_0^4}{L^5}\Big)\frac{M_g^2M_f^2}{M_g^2+M_f^2}\\
&\eta[g]_\phi=\Big(\frac{r_0^3}{L^3}\Big)\frac{M_g^2(M_g^2+\alpha M_f^2)}{M_g^2+M_f^2} \, , \qquad  \eta[g]_\psi=\Big(\frac{r_0^3}{L^3}\Big)\frac{M_g^2M_f^2(1-\alpha)}{M_g^2+M_f^2}.
\end{align}
$P[g]_{\phi}$ represents the pressuer on g-boundary generated by the fluctuation $\phi$. 
We note that the taotal pressure is $P[g]_{\phi}+P[g]_{\psi}=\frac{r_0^4}{L^5}M_g^2$ which is compatible with the value calculated from the background metric Eq.(\ref{bgp}). 
We remember that the entropy density on g-boundary is $s[g]=4\pi M_g^2(r_0/L)^3$ and calculate the ratios 
\begin{align}
\frac{\eta[g]_{\phi}}{s[g]}=\Big(\frac{1}{4\pi}\Big)\frac{M_g^2+\alpha M_f^2}{M_g^2+M_f^2}\,  , \qquad \frac{\eta[g]_{\psi}}{s[g]}=\Big(\frac{1}{4\pi}\Big)\frac{M_f^2(1-\alpha)}{M_g^2+M_f^2}.
\end{align}
If we set $M_g=M_f$, we have the values dependent only on the mass of a graviton  
\begin{align}
\frac{\eta[g]_{\phi}}{s[g]} =\Big(\frac{1}{4\pi}\Big)\frac{1+\alpha}{2}\,  , \qquad \frac{\eta[g]_{\psi}}{s[g]}=\Big(\frac{1}{4\pi}\Big) \frac{1-\alpha}{2}.
\end{align}

\section{conclusion}\label{conclusion}
In this paper, we applied the AdS/CFT correspondence to bimetric gravity in the first order hydrodynamic limit. 
We first reviewed the standard case of pure general relativity, where how to interpret the result is well-known. 
The counterpart on the CFT side is interpreted as fluid of the Yang-Mills plasma. 
Then, we applied this method to dRGT massive gravity and saw that additional divergences emerge. 
In order to cancel these divergences, we added an new counterterm and also imposed a condition on mass of a graviton. 
Though we removed the divergences, how to interpret the result was not clear. 
The AdS/CFT correspondence suggested that the pressure is zero, which contardicts the value calculated from the background metric. 
Thus, we further extended the AdS/CFT prescription to bimetric gravity and expected to remedy the peculiarity. 
As a result, we found that two-component fluid emerge and the total pressure is consistent with the background value. 
We also calculated their sheer viscosity, whici is dependent on the mass of a graviton. \par
However, what we studied in this paper is only the simplest setting. Further detailed investigation is needed to clarify the features of the boundary field theory. 
It is worth studying more complicated perturbations. For example, diagonal perturbations which leads to other properties such as sound waves \cite{Policastro:2002tn} 
or perturbations on the different background $\bar{g}\neq \bar{f}$. 
It may be also interesting to study beyond the first order hydorodynamic limit \cite{Baier:2007ix}. 
The problem of the stability anlysis is left, too. 
In addition, the relation between bimetric or multimetric gravity and the deformation of boundary CFTs \cite{Kiritsis:2006hy, Aharony:2006hz} remains unclear. 
These are left as future works. 
\appendix
\section{counterterm in massive gravity}
In this appendix, we reconsider the counterterm which we introduced in the case of massive gravity Eq.(\ref{Smct}, \ref{SmctPF}). 
Throughout the main part, we take only one type of perturbations such as Eq.(\ref{spur}). 
Then, it is worth considering whether the counterterm Eq.(\ref{Smct}) or Eq.(\ref{SmctPF}) can cancel divergences in other types of perturbations. 
Remembering the calculation of massive gravity, we note that the role of the counterterm Eq.(\ref{SmctPF}) is to cancel the divergence 
coming from the term $u^{-4\alpha}$. This term is a leading order contribution in the expansion of $h=1-u^4$ around $u\sim0$ in Eq.(\ref{metAdSBH}). 
The next order terms continue as $u^{4-4\alpha}$, $u^{8-4\alpha}$,... and so on. 
If we consider pure AdS space-time as a background, we put $h=1$, where only $u^{-4\alpha}$ term is left and we do not have next order divergent terms $O[u^{4-4\alpha}]$. 
Hence, we need only to consider AdS space-time for our purpose. In the following, we set $16\pi G_5=1$ and $L=1$ for notational simplicity. \par
We consider a perturbation dependent only on the variable $u$ 
\begin{align}
{\delta g^{\mu}}_{\nu}=\left(
\begin{array}{ccccc}
\chi_0(u) & -\theta_1(u) & -\theta_2(u) & -\theta_3(u) & -\Pi_0(u) \\
 \theta_1(u)& \chi_1(u)& \phi_1(u)& \phi_2(u)& \Pi_1(u) \\
 \theta_2(u)& \phi_1(u)& \chi_2(u)&\phi_3(u) & \Pi_2(u) \\
 \theta_3(u)&\phi_2(u) &\phi_3(u) & \chi_3(u)& \Pi_3(u) \\
 \Pi_0(u)& \Pi_1(u)& \Pi_2(u)& \Pi_3(u)& \chi_4(u)
\end{array}
\right), 
\end{align}
where the background metric is purely AdS, and minus signs are put to make $\delta g_{\mu \nu}$ symmetric. \par
Now, we expand the action as in section \ref{massivegravity}, and obtain the equations of motion and the on-shell action. 
We skip the details of the calculation, but we find only diagonal terms $\chi$ couple. 
The simplest EOMs come from $\Pi_{i=0,1,2,3}$. They are given by 
\begin{align}
m^2\frac{1}{u^5}\Pi_0=0 \; , \; -m^2\frac{1}{u^5}\Pi_i=0 \: (i=1,\, 2,\, 3).
\end{align}
Their solutions are merely $\Pi_{i=0,1,2,3}=0$. 
The calculation of $\phi_{i=1,2,3}$ is the same as that in section \ref{massivegravity}, so we ommit this part. 
We obtain the EOMs of $\theta_{i=1,2,3}$ as 
\begin{align}
-\frac{1}{u^3}\theta_i''+\frac{3}{u^4}\theta_i'+\frac{m^2}{u^5}\theta_i=0, 
\end{align}
and the solutions are given by 
\begin{align}
\theta_i(u)=A_i u^{2-2\alpha}+B_i u^{2+2\alpha},
\end{align}
where $\alpha=\sqrt{1+m^2/4}$. 
On the other hand, the action of $\theta_i$ is,
\begin{align}
&S=S_{bulk}+S_{GH}+S_{ct}=-\frac{1}{2u^3}\theta_i \theta'_i\Big|_{u=0}\\
&S_{mct}=(1-\alpha)\frac{1}{u^4}\theta_i^2\Big|_{u=0}.
\end{align}
Then, the divergent part of the on-shell action is 
\begin{align}
&S=-A_i^2(1-\alpha)u^{-4\alpha}\\
&S_{mct}=(1-\alpha)A_i^2u^{-4\alpha},
\end{align}
and can be canceled. \par
The $\chi_{i=0,1,2,3,4}$ part is rather complicated. The EOMs are given by 
\begin{align}
&(12+m^2)\chi_4-3u\chi_4'+m^2(\Gamma-\chi_i)+3u(\Gamma'-\chi_i')-u^2(\Gamma''-\chi_i'')=0\quad (i=0,1,2,3) \label{EOMchii} \\
&12\chi_4+m^2\Gamma+3u\Gamma'=0, \label{EOMchi4}
\end{align}
where we put $\Gamma=\chi_0+\chi_1+\chi_2+\chi_3$. Summation of Eq.(\ref{EOMchii}) for $i=0,1,2,3$ is 
\begin{align}
4(12+m^2)\chi_4-12u\chi_4'+3m^2\Gamma+9u\Gamma'-3u^2\Gamma''=0 \label{SUMchii}, 
\end{align}
and we obtain from Eq.(\ref{EOMchi4}) and Eq.(\ref{SUMchii})
\begin{align}
\Gamma=0,\quad \chi_4=0.
\end{align}
Then, EOMs Eq.(\ref{EOMchii}) are written as 
\begin{align}
-m^2\chi_i-3u\chi_i'+u^2\chi_i''=0\quad (i=0,1,2,3)
\end{align}
and we have the solutions 
\begin{align}
&\chi_i(u)=A_iu^{2-2\alpha}+B_iu^{2+2\alpha},\quad (i=1,2,3)\\
&\chi_0(u)=-(A_1+A_2+A_3)u^{2-2\alpha}-(B_1+B_2+B_3)u^{2+2\alpha}.
\end{align}
The action of $\chi$ part is, using $\Gamma=0$,
\begin{align}
&S=S_{bulk}+S_{GH}+S_{ct}=\frac{1}{4u^4}(\chi_0\chi_0'+\chi_1\chi_1'+\chi_2\chi_2'+\chi_3\chi_3')\\
&S_{mct}=(1-\alpha)\frac{1}{u^4}(-\chi_0\chi_0+\chi_1\chi_2+\chi_2\chi_3+\chi_3\chi_1),
\end{align} 
and inserting the solutions of the EOMs, we obtain the divergent part 
\begin{align}
&S=(1-\alpha)(A_1^2+A_2^2+A_3^2+A_1A_2+A_2A_3+A_3A_1)u^{-4\alpha}\Big|_{u=0}\\
&S_{mct}=-(1-\alpha)(A_1^2+A_2^2+A_3^2+A_1A_2+A_2A_3+A_3A_1)u^{-4\alpha}\Big|_{u=0}.
\end{align} 
Thus, all divergence can be canceled. \par
In linear massive gravity, the Pauli-Fiertz mass term not only makes griviton massive but also removes an extra ghost degree of freedom, 
which leads to the transverse traceless condition $\delta {g^{\mu}}_{\mu}=0$, $\nabla_{\mu}\delta {g^{\mu}}_{\nu}=0$. 
Here, these conditions are written as 
\begin{align}
&\chi_0+\chi_1+\chi_2+\chi_3+\chi_4=0\\
&\Pi_i'-\frac{5}{u}\Pi_i=0\, ,\; (i=0,1,2,3)\\
&\frac{1}{u}(-4\chi_4+u\chi_4'+\chi_0+\chi_1+\chi_2+\chi_3)=0,
\end{align}
and compatible with the solutions we obtained. 
\begin{acknowledgments}
I would like to thank Jiro Soda for fruitful discussions. 
K.N. is supported by the Japan Society for the Promotion of Science (JSPS) grant No. 24 -1693.
\end{acknowledgments}


\begin{thebibliography}{27}%
\makeatletter
\providecommand \@ifxundefined [1]{%
 \@ifx{#1\undefined}
}%
\providecommand \@ifnum [1]{%
 \ifnum #1\expandafter \@firstoftwo
 \else \expandafter \@secondoftwo
 \fi
}%
\providecommand \@ifx [1]{%
 \ifx #1\expandafter \@firstoftwo
 \else \expandafter \@secondoftwo
 \fi
}%
\providecommand \natexlab [1]{#1}%
\providecommand \enquote  [1]{``#1''}%
\providecommand \bibnamefont  [1]{#1}%
\providecommand \bibfnamefont [1]{#1}%
\providecommand \citenamefont [1]{#1}%
\providecommand \href@noop [0]{\@secondoftwo}%
\providecommand \href [0]{\begingroup \@sanitize@url \@href}%
\providecommand \@href[1]{\@@startlink{#1}\@@href}%
\providecommand \@@href[1]{\endgroup#1\@@endlink}%
\providecommand \@sanitize@url [0]{\catcode `\\12\catcode `\$12\catcode
  `\&12\catcode `\#12\catcode `\^12\catcode `\_12\catcode `\%12\relax}%
\providecommand \@@startlink[1]{}%
\providecommand \@@endlink[0]{}%
\providecommand \url  [0]{\begingroup\@sanitize@url \@url }%
\providecommand \@url [1]{\endgroup\@href {#1}{\urlprefix }}%
\providecommand \urlprefix  [0]{URL }%
\providecommand \Eprint [0]{\href }%
\providecommand \doibase [0]{http://dx.doi.org/}%
\providecommand \selectlanguage [0]{\@gobble}%
\providecommand \bibinfo  [0]{\@secondoftwo}%
\providecommand \bibfield  [0]{\@secondoftwo}%
\providecommand \translation [1]{[#1]}%
\providecommand \BibitemOpen [0]{}%
\providecommand \bibitemStop [0]{}%
\providecommand \bibitemNoStop [0]{.\EOS\space}%
\providecommand \EOS [0]{\spacefactor3000\relax}%
\providecommand \BibitemShut  [1]{\csname bibitem#1\endcsname}%
\let\auto@bib@innerbib\@empty
\bibitem [{\citenamefont {Maldacena}(1998)}]{Maldacena:1997re}%
  \BibitemOpen
  \bibfield  {author} {\bibinfo {author} {\bibfnamefont {J.~M.}\ \bibnamefont
  {Maldacena}},\ }\href@noop {} {\bibfield  {journal} {\bibinfo  {journal}
  {Adv.Theor.Math.Phys.}\ }\textbf {\bibinfo {volume} {2}},\ \bibinfo {pages}
  {231} (\bibinfo {year} {1998})},\ \Eprint
  {http://arxiv.org/abs/hep-th/9711200} {arXiv:hep-th/9711200 [hep-th]}
  \BibitemShut {NoStop}%
\bibitem [{\citenamefont {Gubser}\ \emph {et~al.}(1998)\citenamefont {Gubser},
  \citenamefont {Klebanov},\ and\ \citenamefont {Polyakov}}]{Gubser:1998bc}%
  \BibitemOpen
  \bibfield  {author} {\bibinfo {author} {\bibfnamefont {S.}~\bibnamefont
  {Gubser}}, \bibinfo {author} {\bibfnamefont {I.~R.}\ \bibnamefont
  {Klebanov}}, \ and\ \bibinfo {author} {\bibfnamefont {A.~M.}\ \bibnamefont
  {Polyakov}},\ }\href {\doibase 10.1016/S0370-2693(98)00377-3} {\bibfield
  {journal} {\bibinfo  {journal} {Phys.Lett.}\ }\textbf {\bibinfo {volume}
  {B428}},\ \bibinfo {pages} {105} (\bibinfo {year} {1998})},\ \Eprint
  {http://arxiv.org/abs/hep-th/9802109} {arXiv:hep-th/9802109 [hep-th]}
  \BibitemShut {NoStop}%
\bibitem [{\citenamefont {Witten}(1998)}]{Witten:1998qj}%
  \BibitemOpen
  \bibfield  {author} {\bibinfo {author} {\bibfnamefont {E.}~\bibnamefont
  {Witten}},\ }\href@noop {} {\bibfield  {journal} {\bibinfo  {journal}
  {Adv.Theor.Math.Phys.}\ }\textbf {\bibinfo {volume} {2}},\ \bibinfo {pages}
  {253} (\bibinfo {year} {1998})},\ \Eprint
  {http://arxiv.org/abs/hep-th/9802150} {arXiv:hep-th/9802150 [hep-th]}
  \BibitemShut {NoStop}%
\bibitem [{\citenamefont {Aharony}\ \emph {et~al.}(2000)\citenamefont
  {Aharony}, \citenamefont {Gubser}, \citenamefont {Maldacena}, \citenamefont
  {Ooguri},\ and\ \citenamefont {Oz}}]{Aharony:1999ti}%
  \BibitemOpen
  \bibfield  {author} {\bibinfo {author} {\bibfnamefont {O.}~\bibnamefont
  {Aharony}}, \bibinfo {author} {\bibfnamefont {S.~S.}\ \bibnamefont {Gubser}},
  \bibinfo {author} {\bibfnamefont {J.~M.}\ \bibnamefont {Maldacena}}, \bibinfo
  {author} {\bibfnamefont {H.}~\bibnamefont {Ooguri}}, \ and\ \bibinfo {author}
  {\bibfnamefont {Y.}~\bibnamefont {Oz}},\ }\href {\doibase
  10.1016/S0370-1573(99)00083-6} {\bibfield  {journal} {\bibinfo  {journal}
  {Phys.Rept.}\ }\textbf {\bibinfo {volume} {323}},\ \bibinfo {pages} {183}
  (\bibinfo {year} {2000})},\ \Eprint {http://arxiv.org/abs/hep-th/9905111}
  {arXiv:hep-th/9905111 [hep-th]} \BibitemShut {NoStop}%
\bibitem [{\citenamefont {Policastro}\ \emph {et~al.}(2001)\citenamefont
  {Policastro}, \citenamefont {Son},\ and\ \citenamefont
  {Starinets}}]{Policastro:2001yc}%
  \BibitemOpen
  \bibfield  {author} {\bibinfo {author} {\bibfnamefont {G.}~\bibnamefont
  {Policastro}}, \bibinfo {author} {\bibfnamefont {D.~T.}\ \bibnamefont {Son}},
  \ and\ \bibinfo {author} {\bibfnamefont {A.~O.}\ \bibnamefont {Starinets}},\
  }\href {\doibase 10.1103/PhysRevLett.87.081601} {\bibfield  {journal}
  {\bibinfo  {journal} {Phys.Rev.Lett.}\ }\textbf {\bibinfo {volume} {87}},\
  \bibinfo {pages} {081601} (\bibinfo {year} {2001})},\ \Eprint
  {http://arxiv.org/abs/hep-th/0104066} {arXiv:hep-th/0104066 [hep-th]}
  \BibitemShut {NoStop}%
\bibitem [{\citenamefont {Horowitz}(2011)}]{Horowitz:2010gk}%
  \BibitemOpen
  \bibfield  {author} {\bibinfo {author} {\bibfnamefont {G.~T.}\ \bibnamefont
  {Horowitz}},\ }\href {\doibase 10.1007/978-3-642-04864-7_10} {\bibfield
  {journal} {\bibinfo  {journal} {Lect.Notes Phys.}\ }\textbf {\bibinfo
  {volume} {828}},\ \bibinfo {pages} {313} (\bibinfo {year} {2011})},\ \Eprint
  {http://arxiv.org/abs/1002.1722} {arXiv:1002.1722 [hep-th]} \BibitemShut
  {NoStop}%
\bibitem [{\citenamefont {Iqbal}\ \emph {et~al.}(2011)\citenamefont {Iqbal},
  \citenamefont {Liu},\ and\ \citenamefont {Mezei}}]{Iqbal:2011ae}%
  \BibitemOpen
  \bibfield  {author} {\bibinfo {author} {\bibfnamefont {N.}~\bibnamefont
  {Iqbal}}, \bibinfo {author} {\bibfnamefont {H.}~\bibnamefont {Liu}}, \ and\
  \bibinfo {author} {\bibfnamefont {M.}~\bibnamefont {Mezei}},\ }\href@noop {}
  {\ ,\ \bibinfo {pages} {707} (\bibinfo {year} {2011})},\ \Eprint
  {http://arxiv.org/abs/1110.3814} {arXiv:1110.3814 [hep-th]} \BibitemShut
  {NoStop}%
\bibitem [{\citenamefont {Kiritsis}(2006)}]{Kiritsis:2006hy}%
  \BibitemOpen
  \bibfield  {author} {\bibinfo {author} {\bibfnamefont {E.}~\bibnamefont
  {Kiritsis}},\ }\href {\doibase 10.1088/1126-6708/2006/11/049} {\bibfield
  {journal} {\bibinfo  {journal} {JHEP}\ }\textbf {\bibinfo {volume} {0611}},\
  \bibinfo {pages} {049} (\bibinfo {year} {2006})},\ \Eprint
  {http://arxiv.org/abs/hep-th/0608088} {arXiv:hep-th/0608088 [hep-th]}
  \BibitemShut {NoStop}%
\bibitem [{\citenamefont {Aharony}\ \emph {et~al.}(2006)\citenamefont
  {Aharony}, \citenamefont {Clark},\ and\ \citenamefont
  {Karch}}]{Aharony:2006hz}%
  \BibitemOpen
  \bibfield  {author} {\bibinfo {author} {\bibfnamefont {O.}~\bibnamefont
  {Aharony}}, \bibinfo {author} {\bibfnamefont {A.~B.}\ \bibnamefont {Clark}},
  \ and\ \bibinfo {author} {\bibfnamefont {A.}~\bibnamefont {Karch}},\ }\href
  {\doibase 10.1103/PhysRevD.74.086006} {\bibfield  {journal} {\bibinfo
  {journal} {Phys.Rev.}\ }\textbf {\bibinfo {volume} {D74}},\ \bibinfo {pages}
  {086006} (\bibinfo {year} {2006})},\ \Eprint
  {http://arxiv.org/abs/hep-th/0608089} {arXiv:hep-th/0608089 [hep-th]}
  \BibitemShut {NoStop}%
\bibitem [{\citenamefont {Apolo}\ and\ \citenamefont
  {Porrati}(2012)}]{Apolo:2012gg}%
  \BibitemOpen
  \bibfield  {author} {\bibinfo {author} {\bibfnamefont {L.}~\bibnamefont
  {Apolo}}\ and\ \bibinfo {author} {\bibfnamefont {M.}~\bibnamefont
  {Porrati}},\ }\href {\doibase 10.1016/j.physletb.2012.07.001} {\bibfield
  {journal} {\bibinfo  {journal} {Phys.Lett.}\ }\textbf {\bibinfo {volume}
  {B714}},\ \bibinfo {pages} {309} (\bibinfo {year} {2012})},\ \Eprint
  {http://arxiv.org/abs/1205.4956} {arXiv:1205.4956 [hep-th]} \BibitemShut
  {NoStop}%
\bibitem [{\citenamefont {Hassan}\ \emph {et~al.}(2012)\citenamefont {Hassan},
  \citenamefont {Rosen},\ and\ \citenamefont {Schmidt-May}}]{Hassan:2011tf}%
  \BibitemOpen
  \bibfield  {author} {\bibinfo {author} {\bibfnamefont {S.}~\bibnamefont
  {Hassan}}, \bibinfo {author} {\bibfnamefont {R.~A.}\ \bibnamefont {Rosen}}, \
  and\ \bibinfo {author} {\bibfnamefont {A.}~\bibnamefont {Schmidt-May}},\
  }\href {\doibase 10.1007/JHEP02(2012)026} {\bibfield  {journal} {\bibinfo
  {journal} {JHEP}\ }\textbf {\bibinfo {volume} {1202}},\ \bibinfo {pages}
  {026} (\bibinfo {year} {2012})},\ \Eprint {http://arxiv.org/abs/1109.3230}
  {arXiv:1109.3230 [hep-th]} \BibitemShut {NoStop}%
\bibitem [{\citenamefont {Hassan}\ and\ \citenamefont
  {Rosen}(2012{\natexlab{a}})}]{Hassan:2011ea}%
  \BibitemOpen
  \bibfield  {author} {\bibinfo {author} {\bibfnamefont {S.}~\bibnamefont
  {Hassan}}\ and\ \bibinfo {author} {\bibfnamefont {R.~A.}\ \bibnamefont
  {Rosen}},\ }\href {\doibase 10.1007/JHEP04(2012)123} {\bibfield  {journal}
  {\bibinfo  {journal} {JHEP}\ }\textbf {\bibinfo {volume} {1204}},\ \bibinfo
  {pages} {123} (\bibinfo {year} {2012}{\natexlab{a}})},\ \Eprint
  {http://arxiv.org/abs/1111.2070} {arXiv:1111.2070 [hep-th]} \BibitemShut
  {NoStop}%
\bibitem [{\citenamefont {Hassan}\ and\ \citenamefont
  {Rosen}(2012{\natexlab{b}})}]{Hassan:2011zd}%
  \BibitemOpen
  \bibfield  {author} {\bibinfo {author} {\bibfnamefont {S.}~\bibnamefont
  {Hassan}}\ and\ \bibinfo {author} {\bibfnamefont {R.~A.}\ \bibnamefont
  {Rosen}},\ }\href {\doibase 10.1007/JHEP02(2012)126} {\bibfield  {journal}
  {\bibinfo  {journal} {JHEP}\ }\textbf {\bibinfo {volume} {1202}},\ \bibinfo
  {pages} {126} (\bibinfo {year} {2012}{\natexlab{b}})},\ \Eprint
  {http://arxiv.org/abs/1109.3515} {arXiv:1109.3515 [hep-th]} \BibitemShut
  {NoStop}%
\bibitem [{\citenamefont {Hinterbichler}\ and\ \citenamefont
  {Rosen}(2012)}]{Hinterbichler:2012cn}%
  \BibitemOpen
  \bibfield  {author} {\bibinfo {author} {\bibfnamefont {K.}~\bibnamefont
  {Hinterbichler}}\ and\ \bibinfo {author} {\bibfnamefont {R.~A.}\ \bibnamefont
  {Rosen}},\ }\href {\doibase 10.1007/JHEP07(2012)047} {\bibfield  {journal}
  {\bibinfo  {journal} {JHEP}\ }\textbf {\bibinfo {volume} {1207}},\ \bibinfo
  {pages} {047} (\bibinfo {year} {2012})},\ \Eprint
  {http://arxiv.org/abs/1203.5783} {arXiv:1203.5783 [hep-th]} \BibitemShut
  {NoStop}%
\bibitem [{\citenamefont {Nomura}\ and\ \citenamefont
  {Soda}(2012)}]{Nomura:2012xr}%
  \BibitemOpen
  \bibfield  {author} {\bibinfo {author} {\bibfnamefont {K.}~\bibnamefont
  {Nomura}}\ and\ \bibinfo {author} {\bibfnamefont {J.}~\bibnamefont {Soda}},\
  }\href {\doibase 10.1103/PhysRevD.86.084052} {\bibfield  {journal} {\bibinfo
  {journal} {Phys.Rev.}\ }\textbf {\bibinfo {volume} {D86}},\ \bibinfo {pages}
  {084052} (\bibinfo {year} {2012})},\ \Eprint {http://arxiv.org/abs/1207.3637}
  {arXiv:1207.3637 [hep-th]} \BibitemShut {NoStop}%
\bibitem [{\citenamefont {Policastro}\ \emph
  {et~al.}(2002{\natexlab{a}})\citenamefont {Policastro}, \citenamefont {Son},\
  and\ \citenamefont {Starinets}}]{Policastro:2002se}%
  \BibitemOpen
  \bibfield  {author} {\bibinfo {author} {\bibfnamefont {G.}~\bibnamefont
  {Policastro}}, \bibinfo {author} {\bibfnamefont {D.~T.}\ \bibnamefont {Son}},
  \ and\ \bibinfo {author} {\bibfnamefont {A.~O.}\ \bibnamefont {Starinets}},\
  }\href {\doibase 10.1088/1126-6708/2002/09/043} {\bibfield  {journal}
  {\bibinfo  {journal} {JHEP}\ }\textbf {\bibinfo {volume} {0209}},\ \bibinfo
  {pages} {043} (\bibinfo {year} {2002}{\natexlab{a}})},\ \Eprint
  {http://arxiv.org/abs/hep-th/0205052} {arXiv:hep-th/0205052 [hep-th]}
  \BibitemShut {NoStop}%
\bibitem [{\citenamefont {Policastro}\ \emph
  {et~al.}(2002{\natexlab{b}})\citenamefont {Policastro}, \citenamefont {Son},\
  and\ \citenamefont {Starinets}}]{Policastro:2002tn}%
  \BibitemOpen
  \bibfield  {author} {\bibinfo {author} {\bibfnamefont {G.}~\bibnamefont
  {Policastro}}, \bibinfo {author} {\bibfnamefont {D.~T.}\ \bibnamefont {Son}},
  \ and\ \bibinfo {author} {\bibfnamefont {A.~O.}\ \bibnamefont {Starinets}},\
  }\href {\doibase 10.1088/1126-6708/2002/12/054} {\bibfield  {journal}
  {\bibinfo  {journal} {JHEP}\ }\textbf {\bibinfo {volume} {0212}},\ \bibinfo
  {pages} {054} (\bibinfo {year} {2002}{\natexlab{b}})},\ \Eprint
  {http://arxiv.org/abs/hep-th/0210220} {arXiv:hep-th/0210220 [hep-th]}
  \BibitemShut {NoStop}%
\bibitem [{\citenamefont {de~Rham}\ and\ \citenamefont
  {Gabadadze}(2010)}]{deRham:2010ik}%
  \BibitemOpen
  \bibfield  {author} {\bibinfo {author} {\bibfnamefont {C.}~\bibnamefont
  {de~Rham}}\ and\ \bibinfo {author} {\bibfnamefont {G.}~\bibnamefont
  {Gabadadze}},\ }\href {\doibase 10.1103/PhysRevD.82.044020} {\bibfield
  {journal} {\bibinfo  {journal} {Phys.Rev.}\ }\textbf {\bibinfo {volume}
  {D82}},\ \bibinfo {pages} {044020} (\bibinfo {year} {2010})},\ \Eprint
  {http://arxiv.org/abs/1007.0443} {arXiv:1007.0443 [hep-th]} \BibitemShut
  {NoStop}%
\bibitem [{\citenamefont {de~Rham}\ \emph {et~al.}(2011)\citenamefont
  {de~Rham}, \citenamefont {Gabadadze},\ and\ \citenamefont
  {Tolley}}]{deRham:2010kj}%
  \BibitemOpen
  \bibfield  {author} {\bibinfo {author} {\bibfnamefont {C.}~\bibnamefont
  {de~Rham}}, \bibinfo {author} {\bibfnamefont {G.}~\bibnamefont {Gabadadze}},
  \ and\ \bibinfo {author} {\bibfnamefont {A.~J.}\ \bibnamefont {Tolley}},\
  }\href {\doibase 10.1103/PhysRevLett.106.231101} {\bibfield  {journal}
  {\bibinfo  {journal} {Phys.Rev.Lett.}\ }\textbf {\bibinfo {volume} {106}},\
  \bibinfo {pages} {231101} (\bibinfo {year} {2011})},\ \Eprint
  {http://arxiv.org/abs/1011.1232} {arXiv:1011.1232 [hep-th]} \BibitemShut
  {NoStop}%
\bibitem [{\citenamefont {Hassan}\ and\ \citenamefont
  {Rosen}(2011)}]{Hassan:2011vm}%
  \BibitemOpen
  \bibfield  {author} {\bibinfo {author} {\bibfnamefont {S.}~\bibnamefont
  {Hassan}}\ and\ \bibinfo {author} {\bibfnamefont {R.~A.}\ \bibnamefont
  {Rosen}},\ }\href {\doibase 10.1007/JHEP07(2011)009} {\bibfield  {journal}
  {\bibinfo  {journal} {JHEP}\ }\textbf {\bibinfo {volume} {1107}},\ \bibinfo
  {pages} {009} (\bibinfo {year} {2011})},\ \Eprint
  {http://arxiv.org/abs/1103.6055} {arXiv:1103.6055 [hep-th]} \BibitemShut
  {NoStop}%
\bibitem [{\citenamefont {Son}\ and\ \citenamefont
  {Starinets}(2002)}]{Son:2002sd}%
  \BibitemOpen
  \bibfield  {author} {\bibinfo {author} {\bibfnamefont {D.~T.}\ \bibnamefont
  {Son}}\ and\ \bibinfo {author} {\bibfnamefont {A.~O.}\ \bibnamefont
  {Starinets}},\ }\href {\doibase 10.1088/1126-6708/2002/09/042} {\bibfield
  {journal} {\bibinfo  {journal} {JHEP}\ }\textbf {\bibinfo {volume} {0209}},\
  \bibinfo {pages} {042} (\bibinfo {year} {2002})},\ \Eprint
  {http://arxiv.org/abs/hep-th/0205051} {arXiv:hep-th/0205051 [hep-th]}
  \BibitemShut {NoStop}%
\bibitem [{\citenamefont {Skenderis}(2002)}]{Skenderis:2002wp}%
  \BibitemOpen
  \bibfield  {author} {\bibinfo {author} {\bibfnamefont {K.}~\bibnamefont
  {Skenderis}},\ }\href {\doibase 10.1088/0264-9381/19/22/306} {\bibfield
  {journal} {\bibinfo  {journal} {Class.Quant.Grav.}\ }\textbf {\bibinfo
  {volume} {19}},\ \bibinfo {pages} {5849} (\bibinfo {year} {2002})},\ \Eprint
  {http://arxiv.org/abs/hep-th/0209067} {arXiv:hep-th/0209067 [hep-th]}
  \BibitemShut {NoStop}%
\bibitem [{\citenamefont {Breitenlohner}\ and\ \citenamefont
  {Freedman}(1982{\natexlab{a}})}]{Breitenlohner:1982bm}%
  \BibitemOpen
  \bibfield  {author} {\bibinfo {author} {\bibfnamefont {P.}~\bibnamefont
  {Breitenlohner}}\ and\ \bibinfo {author} {\bibfnamefont {D.~Z.}\ \bibnamefont
  {Freedman}},\ }\href {\doibase 10.1016/0370-2693(82)90643-8} {\bibfield
  {journal} {\bibinfo  {journal} {Phys.Lett.}\ }\textbf {\bibinfo {volume}
  {B115}},\ \bibinfo {pages} {197} (\bibinfo {year}
  {1982}{\natexlab{a}})}\BibitemShut {NoStop}%
\bibitem [{\citenamefont {Breitenlohner}\ and\ \citenamefont
  {Freedman}(1982{\natexlab{b}})}]{Breitenlohner:1982jf}%
  \BibitemOpen
  \bibfield  {author} {\bibinfo {author} {\bibfnamefont {P.}~\bibnamefont
  {Breitenlohner}}\ and\ \bibinfo {author} {\bibfnamefont {D.~Z.}\ \bibnamefont
  {Freedman}},\ }\href {\doibase 10.1016/0003-4916(82)90116-6} {\bibfield
  {journal} {\bibinfo  {journal} {Annals Phys.}\ }\textbf {\bibinfo {volume}
  {144}},\ \bibinfo {pages} {249} (\bibinfo {year}
  {1982}{\natexlab{b}})}\BibitemShut {NoStop}%
\bibitem [{\citenamefont {Mezincescu}\ and\ \citenamefont
  {Townsend}(1985)}]{Mezincescu:1984ev}%
  \BibitemOpen
  \bibfield  {author} {\bibinfo {author} {\bibfnamefont {L.}~\bibnamefont
  {Mezincescu}}\ and\ \bibinfo {author} {\bibfnamefont {P.}~\bibnamefont
  {Townsend}},\ }\href {\doibase 10.1016/0003-4916(85)90150-2} {\bibfield
  {journal} {\bibinfo  {journal} {Annals Phys.}\ }\textbf {\bibinfo {volume}
  {160}},\ \bibinfo {pages} {406} (\bibinfo {year} {1985})}\BibitemShut
  {NoStop}%
\bibitem [{\citenamefont {Baier}\ \emph {et~al.}(2008)\citenamefont {Baier},
  \citenamefont {Romatschke}, \citenamefont {Son}, \citenamefont {Starinets},\
  and\ \citenamefont {Stephanov}}]{Baier:2007ix}%
  \BibitemOpen
  \bibfield  {author} {\bibinfo {author} {\bibfnamefont {R.}~\bibnamefont
  {Baier}}, \bibinfo {author} {\bibfnamefont {P.}~\bibnamefont {Romatschke}},
  \bibinfo {author} {\bibfnamefont {D.~T.}\ \bibnamefont {Son}}, \bibinfo
  {author} {\bibfnamefont {A.~O.}\ \bibnamefont {Starinets}}, \ and\ \bibinfo
  {author} {\bibfnamefont {M.~A.}\ \bibnamefont {Stephanov}},\ }\href {\doibase
  10.1088/1126-6708/2008/04/100} {\bibfield  {journal} {\bibinfo  {journal}
  {JHEP}\ }\textbf {\bibinfo {volume} {0804}},\ \bibinfo {pages} {100}
  (\bibinfo {year} {2008})},\ \Eprint {http://arxiv.org/abs/0712.2451}
  {arXiv:0712.2451 [hep-th]} \BibitemShut {NoStop}%
\bibitem{Natsuume} M. Natsuume, The AdS/CFT Duality User Guide (Springer, to appear)
\end{thebibliography}

%

\end{document}